\documentclass[%
aps,amsmath,amssymb,floatfix, onecolumn, prf
]{revtex4-2}

\usepackage{graphicx}
\usepackage{dcolumn}
\usepackage{bm}

\graphicspath{{./pic/}}
\usepackage{color}
\usepackage{amsmath}
\usepackage{tabularx}
\usepackage{rotating}
\usepackage{epic}
\usepackage{caption}
\begin{document}

\preprint{APS/123-QED}
\title{Fluctuation-driven dynamics in nanoscale thin-film flows: physical insights from numerical investigations  }

\author{Chengxi Zhao}
\email{zhaochengxi@ustc.edu.cn}
\affiliation{Department of Modern Mechanics, University of Science and Technology of China, Hefei 230026, China}
\affiliation{School of Engineering, University of Warwick, Coventry CV4 7AL, United Kingdom 
}
\author{Jingbang Liu}
\email{jingbang.liu@warwick.ac.uk}
\affiliation{
Mathematics Institute, University of Warwick, Coventry CV4 7AL, United Kingdom
}

\author{Duncan A. Lockerby}
 \email{D.Lockerby@warwick.ac.uk}
\affiliation{School of Engineering, University of Warwick, Coventry CV4 7AL, United Kingdom }

\author{James E. Sprittles}
\email{J.E.Sprittles@warwick.ac.uk}
\affiliation{Mathematics Institute, University of Warwick, Coventry CV4 7AL, United Kingdom
}



\begin{abstract}
The effects of thermal fluctuations on nanoscale flows are captured by a numerical scheme that is underpinned by fluctuating hydrodynamics. A stochastic lubrication equation (SLE) is solved on non-uniform adaptive grids to study a series of nanoscale thin-film flows. 
The Fornberg scheme is used for high-resolution spatial discretisation and a fully-implicit time-marching scheme is designed for numerical stability.
The accuracy of the numerical method is verified against theoretical results for thermal capillary waves during the linear stage of their development.
The framework is then used to study the nonlinear behaviour of three bounded thin-film flows: (i) droplet spreading, where new power laws are derived; (ii) droplet coalescence, where molecular dynamics results are reproduced by the SLE at a fraction of the computational cost and it is discovered that thermal fluctuations decelerate the process, in contrast to previously investigated phenomena;  and (iii) thin-film rupture, where, in the regime considered, disjoining pressure dominates the final stages of rupture.
\end{abstract}

\maketitle

\section{INTRODUCTION}

Bounded planar thin-film flows are common in both nature and technology. A variety of interesting flow behaviours fall into this class, including rupture \cite{jacobs1998thin}, dewetting \cite{baumchen2014influence}, droplet spreading \cite{tanner1979spreading} and sessile droplet coalescence \cite{hernandez2012symmetric}. Dynamics of these processes can often be well described by lubrication equations (LE), derived using a long-wave approximation to the Navier-Stokes equations \cite{oron1997long}, which reduces the  modelling problem to solving a single partial differential equation.
Motivated by emerging technologies in micro and nanoscale fluid dynamics \cite{stone2004engineering}, thin-film nanoflows have attracted considerable interest recently and challenged conventional theories, due to the emergence of new dominant physics at these scales.
An important physical factor at the nanoscale is thermal noise, which has been shown experimentally to influence interfacial dynamics through the observation of thermal capillary waves on interfaces of (ultra-low surface-tension) colloid-polymer mixtures \cite{aarts2004direct}.
Recently, by the use of molecular dynamics (MD) simulations, it has been discovered that fluctuation-driven nanowaves dominate a range of nanoscale interfacial-flow phenomena, such as nanothread breakup \cite{moseler2000formation}, nanojet instablity \cite{zhao2019revisiting}, nanodroplet coalescence \cite{perumanath2019droplet} and development of rough interfaces on nanoscale thin films \cite{zhang2021thermal}. 
Notably, the observations of nanothread breakup in MD were further confirmed by an analytical model \cite{eggers2002dynamics} and experiments with colloid-polymer mixtures \cite{hennequin2006drop, petit2012break}.

To model thin-film flows with thermal noise mathematically,  Gr{\"u}n et al. \cite{grun2006thin} derived a stochastic lubrication equation (SLE) by applying a long-wave approximation to the Landau-Lifshitz-Navier-Stokes equations (the fluctuating hydrodynamics equations) \cite{landau1976statistical}. 
Linear instability analysis was then conducted on the SLE by Mecke \& Rauscher \cite{mecke2005thermal} to obtain an evolving spectrum for thermal capillary waves of the film interface. 
It was shown that thermal noise changes the spectrum of thermal capillary waves from an exponential decay to a power law for large wavenumbers, 
which was then confirmed by the experimental observations on the dewetting of polymer films \cite{fetzer2007thermal}.
The same behaviour of spectrum was also confirmed in MD simulations, first done by Willis \& Freund \cite{willis2010thermal} in 2010 and recently, followed by Zhang et al. \cite{zhang2019molecular}, who showed that interfacial roughening falls into a universality class \cite{zhang2021thermal}.  However, theories based on small deviations from equilibrium cannot, inevitably, describe inherently nonlinear events such as film breakup.

Besides the theoretical analysis, numerical studies of the fully nonlinear SLE have also been carried out with uncorrelated  noise \cite{nesic2015fully} and a spatially correlated noise model \cite{diez2016metallic}, whose solutions at the linear stage are confirmed by the theory for thermal capillary waves and experimental data, respectively.
The nonlinear dynamics was also investigated in these numerical studies with particular attention paid to thin-film rupture time \cite{diez2016metallic} and droplet size distribution after dewetting \cite{nesic2015fully}.
However, analytic results for these nonlinear behaviours are non-trivial.
The only successful attempt is on nano-droplet spreading \cite{davidovitch2005spreading, nesic2015dynamics}, where a similarity solution was derived for `fluctuation-dominated spreading' and then verified numerically.
Therefore, numerical investigations have become an important approach to better understand the nonlinear behaviours of thin-film flows. Interestingly, the spatial discretisation in previous works  \cite{nesic2015fully, diez2016metallic,shah2019thermal} was developed only on uniform grids, meaning that the locally nonlinear behaviours (e.g. multiscale rupture dynamics) could not be easily resolved with high accuracy, while, in the deterministic cases (LE), these local behaviours have been captured accurately on non-uniform grids in a range of different numerical schemes \cite{diez2002computing, pahlavan2018thin, zitz2019lattice,lam2019computing}.

In this work, we develop an accurate and efficient numerical scheme (on non-uniform grids) to overcome the aforementioned drawbacks and employ it to investigate different kinds of bounded thin-film flows; in particular for flows where we expect locally nonlinear dynamics (such as in rupture and coalescence).

The article is laid out as follows.
In Sec\,.\ref{sec_model}, the SLE is introduced, the numerical scheme is proposed and the correlated-noise model is presented.
In Sec.\,\ref{sec_ver}, numerical solutions for the SLE are verified against known analytical results (Sec.\,\ref{sec_ver}).
In Sec.\,\ref{sec_fur}, we use the SLE solver to study three different kinds of thin-film flows with nonlinear dynamics: (i) droplet spreading on a precursor film verified by new theories developed (Sec.\,\ref{sec_spreading}); (ii) bounded droplet coalescence, validated by MD (Sec.\,\ref{sec_coal}) and (iii) thin-film rupture (Sec.\,\ref{sec_rup}).

\section{Mathematical Modelling and Numerical Approach} \label{sec_model}

In this section, we introduce the mathematical model which describes nanoscale thin-film flows and then propose a numerical framework for solving the associated system of equations. 
We first present the non-dimensionalised SLE.

\subsection{Stochastic lubrication equations (SLE) \label{sec_SLE}}
\begin{figure}[h]
\centering
\includegraphics[width=0.62\textwidth]{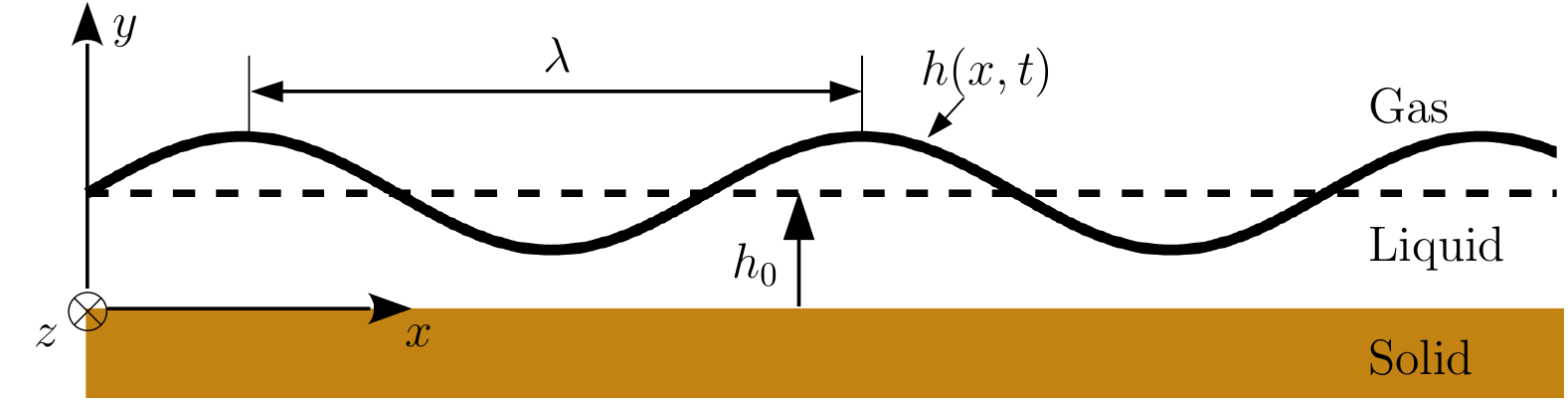}
	\caption{Schematic of a thin film with a perturbed interface  } 
\label{fig_set_up}	
\end{figure}

The SLE for a two-dimensional bounded film was first derived by Gr{\"u}n et al. \cite{grun2006thin} and Davidovitch et al. \cite{davidovitch2005spreading}, who applied a long-wave approximation to the fluctuating hydrodynamics equations to describe the dynamics of the interface by the film height $h(x,t)$ (see Figure\,\ref{fig_set_up}). 
Recently, Zhang et al. \cite{zhang2020nanoscale} proposed a more general SLE using a slip boundary condition, which is crucial at the nanoscale, so this approach is employed here.

To identify the governing dimensionless parameters, we non-dimensionalise the SLE with the characteristic scales
shown below:
$$h = \tilde{h}/h_0, \quad t = \tilde{t}/(3\mu h_0/\gamma), 
\quad \Pi = \tilde{\Pi}/(\gamma/h_0), 
\quad \mathcal{N} = \mathcal{\tilde{N}} \sqrt{3 \mu h_0^2/\gamma},$$
where $\tilde{h}$, $\tilde{t}$, $\tilde{\Pi}$ and $\tilde{\mathcal{N}}$ represent the dimensional interface height, time, disjoining pressure and normally distributed random variable (the model for thermal fluctuations), respectively. Note that the dimensional material parameters are not given tildes.   
$h_0$ is the characteristic film height, $\mu$ is the liquid's dynamic viscosity and $\gamma$ is surface tension.
Following Zhang et al.\cite{zhang2020nanoscale}, the dimensionless SLE is given by:
\begin{equation}
\label{eq_nondim_SLE}
\frac{\partial h}{\partial t} = 
-\frac{\partial}{\partial x}\left[ M(h) \left( \frac{\partial^3 h}{\partial x^3}- \frac{\partial \Pi}{\partial x} \right) - \sqrt{2 \varphi M(h)}\,  \mathcal{N} \right] \,.
\end{equation}
Here, the mobility $M(h) = h^3+3\, l_\mathrm{s} h^2$, so that the slip length $l_\mathrm{s}=0$ the `orginal SLE' from Ref.\,\cite{grun2006thin} is recovered.
The disjoining pressure $\Pi=A / (6 \pi h^3)$, where $A$ is the dimensionless Hamaker constant, reflecting the strength of the van der Waals forces between liquid and a substrate \cite{zhang2019molecular}. 
The noise term $\mathcal{N}$ has zero mean and covariance $\langle \mathcal{N}(x,t) \mathcal{N}(x',t') \rangle = \delta(x-x')\delta(t-t')$.
The dimensionless parameter $\varphi = l^2_\mathrm{T}/(W h_0)$ relates to the intensity of interface fluctuations, where $l_\mathrm{T}= \sqrt{k_\mathrm{B} T/\gamma}$ is the characteristic thermal fluctuation length and $W$ is the initial thickness of the film ($z$-direction).
When $\varphi=0$, the deterministic LE \cite{oron1997long} is recovered.
In this work, periodic boundary conditions are considered in all cases and the initial conditions depend on the type of thin-film flow we simulate.
The remaining parts of this section are concerned with the numerical scheme for solving Equation\,(\ref{eq_nondim_SLE}).

\subsection{Spatial discretisation on non-uniform grids \label{SLE_num}}

To resolve locally large gradients in flow variables, without slowing down the computation dramatically, we use a non-uniform grid for spatial discretisation. We choose the well-known scheme proposed by Fornberg \cite{fornberg1998classroom} to approximate spatial derivatives using finite differences,
\begin{equation}
\frac{d^k}{dx^k}h(x)\approx\sum_{j=0}^{n}\mathcal{L}_{j,n}^k(x) h(x_j),
\end{equation}
where $\{x_j\}^{s}_{j=0}$ are the grid points chosen for approximation and $\mathcal{L}_{j,s}^k(x)$ are the Fornberg coefficients calculated at $x$. Details of the calculation and error estimate of the Fornberg coefficients can be found in Appendix\,\ref{APP_Fornberg}.
 
To apply the Fornberg scheme for spatial discretisation we first rearrange the SLE into a conservative form:
\begin{equation}
\frac{\partial h}{\partial t} = -\frac{\partial F(h)}{\partial x}, \quad \mathrm{where}\,\, F(h) = M(h) \left( \frac{\partial^3 h}{\partial x^3}+ \frac{A}{2\pi h^4}\frac{\partial h}{\partial x} \right) - \sqrt{2 \varphi M(h)}\,  \mathcal{N}.
\label{eq_SLE_num_spatial}
\end{equation}
The film height $h(x,t)$ is to be solved at the $m$-th step, $t^m$, according to a certain spatial discretisation $\{x_i\}_{i=0}^n$ to give an array $\{h_i^m\}_{i=0}^n$.
We use 3 ($n=2$) and 5 ($n=4$) points to approximate the first-order and the third-order derivatives, respectively, with the target point located at the centre to provide second-order accuracy. The expression for the spatial discretisation of equation (\ref{eq_SLE_num_spatial}) at the $i$-th node is then written as:
\begin{equation}
\partial_t h_i  = -\sum^{2}_{j=0} \mathcal{L}^1_{j,2} F_{i-1+j},
\label{eq_SLE_num_spatial2}
\end{equation}
where
\begin{align}
F_{i} = M(h_i) \left(
\sum^{4}_{j=0} \mathcal{L}^3_{j,4} h_{i-2+j} + 
\frac{ A}{2 \pi h_i^4}\sum^{2}_{j=0} \mathcal{L}^1_{j,2} h_{i-1+j}\right)
-\sqrt{2 \varphi M(h_i)} N_i \,.
\end{align}
Here, $\mathcal{L}^1_{j,2}(x_i)$ using $\{x_{i-1}, x_i, x_{i+1}\}$ and $\mathcal{L}^3_{j,4}(x_i)$ using $\{x_{i-2}, x_{i-1}, x_i, x_{i+1}, x_{i+2}\}$ are the Fornberg coefficients for the first and third derivative, respectively.

\subsection{Implicit time-marching method}
Equation\,(\ref{eq_SLE_num_spatial2}) is of the form
\begin{equation}
\frac{\partial \mathbf{h}}{\partial t} = \mathbf{D}\,\mathbf{h}, 
\label{eq_SLE_num_temp}
\end{equation}
where $\mathbf{h}$ is the solution vector and $\mathbf{D}(\mathbf{h})$ is a matrix representing the nonlinear differential operator and has size $(n+1) \times (n+1)$.
Employing the implicit Euler time marching method \cite{kloeden1992stochastic} to Equation\,(\ref{eq_SLE_num_temp}) gives
\begin{equation}
\frac{\mathbf{h}^{m+1}-\mathbf{h}^{m}}{\triangle t} = \mathbf{D}( \mathbf{h}^{m+1} ) \,\mathbf{h}^{m+1},
\end{equation}
where the superscript $m$ denotes the time-step level. This is equivalent to
\begin{equation}
\mathbf{G}=\left[ \mathbf{I}-\triangle t\,
 \mathbf{D} ( \mathbf{h}^{m+1} )  \right] \mathbf{h}^{m+1} - \mathbf{h}^{m}=\mathbf{0} \,,
\label{eq_SLE_num_temp2}
\end{equation}
and this root-finding problem, $\mathbf{G}(\mathbf{h}^{m+1})=\mathbf{0}$, can be solved by the Newton-Kantorovich method \cite{polyak2006newton}, using an initial guess of the solution $\mathbf{h}^{g}$ (the superscript $g$ denotes `guess') obtained by an explicit Euler time marching method:
\begin{equation}
	\label{eq_initial_guess}
	\mathbf{h}^g = \mathbf{h}^m+ \triangle t\, \mathbf{D}(\mathbf{h}^m)\, \mathbf{h}^m .
\end{equation}
If $\mathbf{h}^g$ is a distance $\mathbf{q}$ away from the solution, i.e., $\mathbf{h}^{m+1} = \mathbf{h}^g+\mathbf{q}$, so that $\mathbf{G}(\mathbf{h}^g+\mathbf{q})=\mathbf{0}$, then in the linear approximation one has 
\begin{equation}
\label{eq_newton1}
\mathbf{G}(\mathbf{h}^g+\mathbf{q}) \approx \mathbf{G}(\mathbf{h}^g) + \frac{\partial \mathbf{G}(\mathbf{h}^g)}{\partial \mathbf{h}^g} \mathbf{q} = \mathbf{0} \,.
\end{equation}
From Equation\,(\ref{eq_SLE_num_temp2}) we have
\begin{align}
\label{eq_newton2}
\frac{\partial \mathbf{G}(\mathbf{h}^g)}{\partial \mathbf{h}^g}  
 = \mathbf{I} - \triangle t \mathbf{J} \,,
\end{align}
where the $\mathbf{J}$ is the Jacobian of $\mathbf{D} (\mathbf{h}^g)$. Combining
Equation\,(\ref{eq_newton1}) and (\ref{eq_newton2}) yields a linear approximation of $\mathbf{q}$,
\begin{equation}
	\mathbf{q} \approx -(\mathbf{I}-\triangle t\mathbf{J})^{-1}\mathbf{G}(\mathbf{h}^g),
\end{equation}
so that we can update $\mathbf{h}^g$ as $\mathbf{h}^g+\mathbf{q}$ and repeat the process. If the iteration
converges, both $\mathbf{q}$ and $\mathbf{G}(\mathbf{h}^g) $ will decrease quadratically, eventually to zero.
The final solution, $\mathbf{h}^{m+1} = \mathbf{h}^{g}$.

To control numerical errors caused by the temporal discretisation, we implement the following criteria:
\begin{itemize}
	\item The new solution is not negative at any point, i.e., $h_i > 0$;
	\item The maximum value of the time derivative error has to be smaller than a prescribed upper limit, where the error can be computed using the second-order derivative in the Taylor expansion;
	i.e., $\max \limits_{1\leq i \leq n} \left( \left| \frac{(\triangle t^m)^2}{h^m_i} \frac{d^2 h^m_i}{dt^2} \right| \right) <10^{-3}$;
	\item The iteration process converges, i.e., specifically,  $|q_i|/h_i$ decreases monotonically to below $ 10^{-4}$;
	\item The number of iteration steps is smaller than 100. 
\end{itemize}

If any condition is not satisfied, we go back to the initial value
($h^g_i = h^m_i$) and restart the iteration process with a reduced
time step. If the new $\triangle t$ is smaller than $10^{-16}$, the iteration halts and the code ends in a failed state.
Unavoidably, the implicit Euler Scheme applied to stochastic differential equations converges slower with reducing time steps than the equivalent deterministic system \cite{kloeden1992stochastic}, forcing a smaller time step for the SLE (compared to the LE) that significantly increases the computational costs.

\subsection{Correlated-noise model}

The covariance of uncorrelated fluctuations are described  by a Dirac delta function.
The delta function could be approximated by a 2D rectangular (boxcar) function (in $t$ and $x$) that is non-zero over a time step ($\triangle t$) and grid spacing ($\triangle x$), i.e., $\mathcal{N}(x,t) \approx N_i^t / \sqrt{\triangle t \triangle x}$.
Here, $N_i^t$ represents computer-generated random numbers, that follows a normal distribution with zero mean and unit variance.  
However, this model has been shown prone to numerical instability for the SLE \cite{zhao2020dynamics}; problems that are exacerbated as $\triangle x$ and $\triangle t$ become smaller and the amplitude of noise becomes larger.

To achieve a robust numerical scheme with non-uniform spatial and temporal discretisation, we combine the spatially correlated noise model from Ref.\,\cite{grun2006thin} and the temporal one from Ref.\,\cite{zhao2020dynamics}, so that the noise becomes correlated beneath the spatial correlation length $L_\mathrm{c}$ and the time correlation length $T_\mathrm{c}$.
Uncorrelated behaviour can then be approximated by taking the limit of these lengths to zero, ensuring they are numerically well resolved throughout the limiting process.

Following \cite{grun2006thin}, the stochastic term, $\mathcal{N}(x,t)$ is expanded using separation of variables in the Q-Wiener ($\mathcal{W}(x,t)$) process as,
\begin{equation}
\mathcal{N}(x,t) = \frac{\partial \mathcal{W}(x,t)}{\partial t} 
= \sum^{q \rightarrow +\infty}_{q \rightarrow -\infty} \chi_q\, \dot{c}_q(t)\, g_q(x) \,.
\end{equation}
Here, the constant $\chi_q$ are the eigenvalues of the correlation function $F_\mathrm{cor}$,
\begin{equation}
\chi_q = \int_{-L/2}^{L/2} F_\mathrm{cor}(x) e^{-i\, 2\pi q z/L} dx\,
\end{equation}
where $q$ represents an integer sequence.
The expressions for $F_\mathrm{cor}$ and details about this model can be found in Appendix~\ref{App_noise_model}.
The coefficient $\dot{c}_q(t)$ represents a temporally correlated noise process in our setup, in contrast to \cite{grun2006thin} where uncorrelated noise is considered, modelled by a simple linear interpolation between uncorrelated random noise at the endpoints of the temporal correlation interval \cite{zhao2020dynamics} (see Figure\,\ref{fig_app_linear}). 
As the full implicit time marching method usually provides stable numerical performance with large time steps ($\triangle t > T_\mathrm{c}$), this temporally correlated model will be only activated, and is only practically necessary, when capturing local dynamics near singularities for very small $\triangle t$, e.g., for the final stage of film rupture (see more details in Appendix~\ref{App_noise_model}).

\subsection{Grid-size convergence}
\begin{figure}[h]
\centering
\captionsetup{justification=raggedright}
\includegraphics[width=0.55\textwidth]{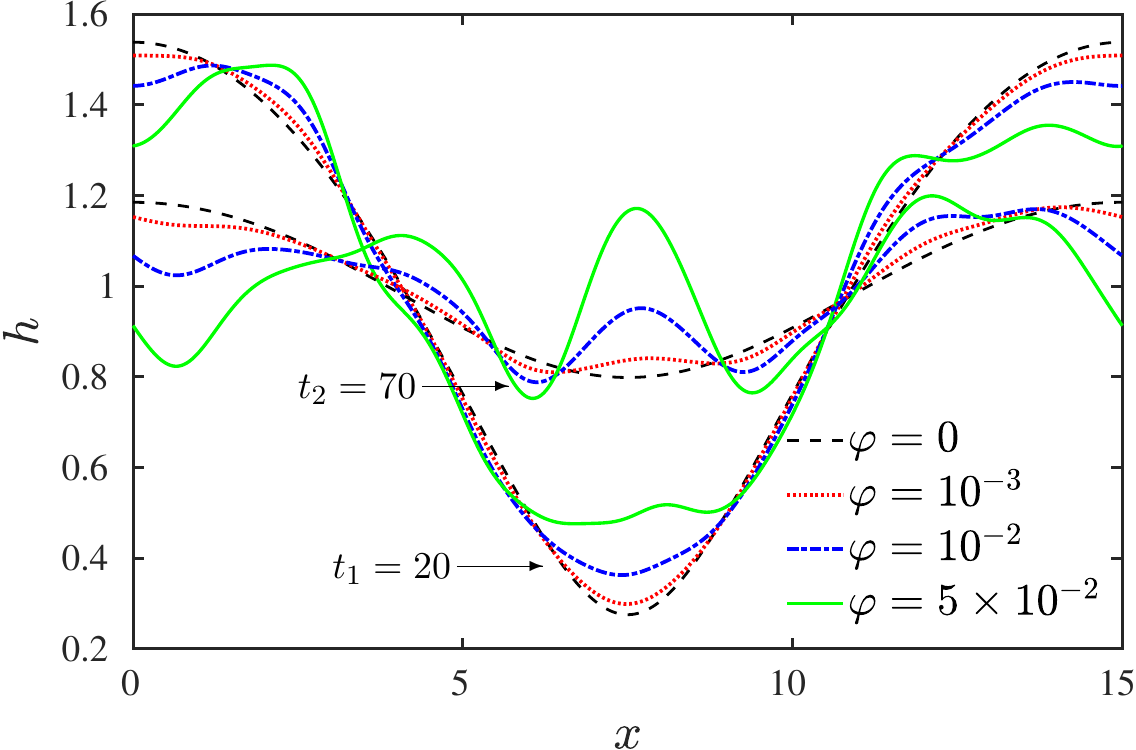}
	\caption{Interface profiles at two time instants ($t_1 = 20$ and $t_2 = 70$) from numerical solutions of the SLE with different values of the noise strength $\varphi$. Note that the same random numbers are employed in all the cases.}
\label{fig_SLE_covergence_pro}	
\end{figure}

To test the grid-size convergence of the numerical approach above, we consider the simulation of a relatively short film $L=15$  with an initial perturbation $h(x,0) = -0.7\,\mathrm{cos}(2\,\pi\,x/L)$, which relaxes to a relatively flat film in the absence of disjoining pressure ($A=0$).
We set $L_\mathrm{c}=1.5$ and run the simulations with an increasingly fine grid spacing for different values of the noise strength $\varphi$, whose influence is shown in Figure\,\ref{fig_SLE_covergence_pro}.

\begin{figure}[h]
\centering
\captionsetup{justification=raggedright}
\includegraphics[width=1.0\textwidth]{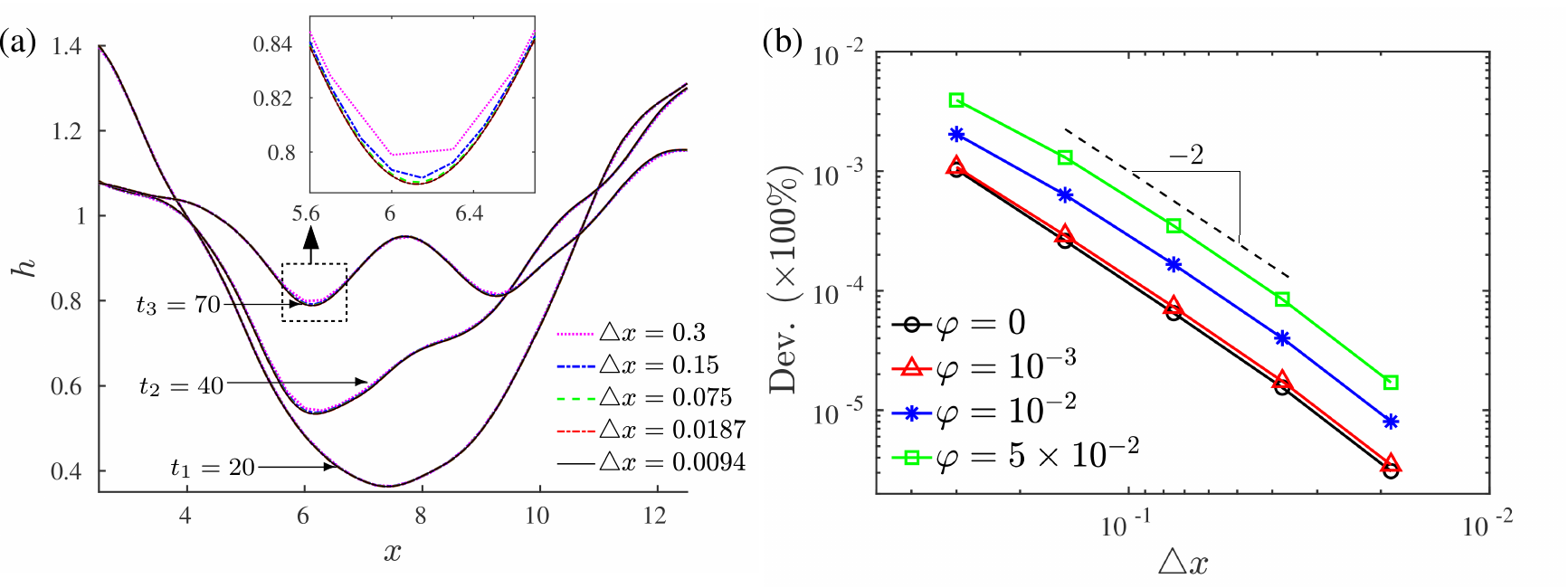}
	\caption{(a) Interface profiles obtained with different grids at three time instances: $t_1 = 20$, $t_2 = 40$ and $t_3 = 70$. Here, $\varphi=10^{-2}$.
	(b) Convergence characteristics for decreasing grid size, where Dev. is average deviation of interface profiles to the finest resolution profile in (a).}
\label{fig_SLE_covergence}	
\end{figure}

\begin{figure}[h]
\centering
\captionsetup{justification=raggedright}
\includegraphics[width=0.99\textwidth]{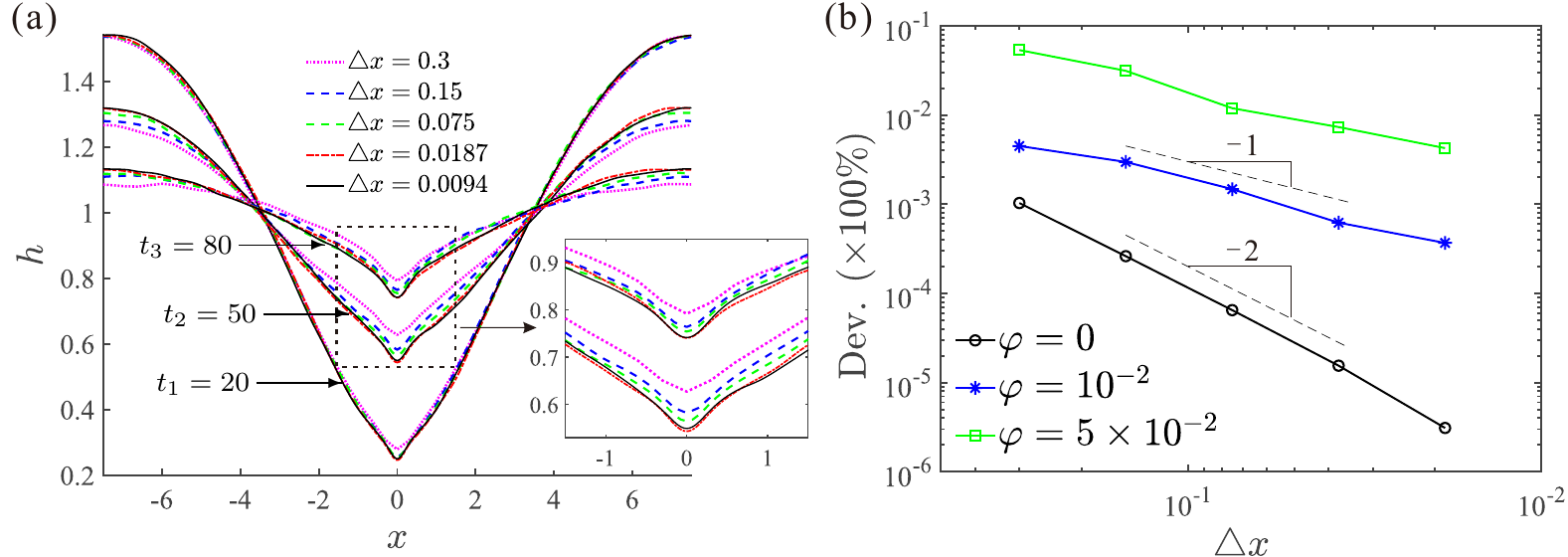}
	\caption{ (a) Ensemble-averaged interface profiles obtained from 200 independent realisations at three time instances: $t_1 = 20$, $t_2 = 50$ and $t_3 = 80$. Here, $\varphi=10^{-2}$.
	(b) Convergence characteristics for decreasing grid size, where Dev. is average deviation of ensemble-averaged interface profiles to the finest resolution profile in (a).} 
\label{fig_SLE_covergence2}	
\end{figure}

Notably, besides numerical errors caused by discretisations in the simulation, statistical errors are also introduced by the stochastic term in the SLE. 
To better demonstrate the convergence, we measure two different kinds of errors.

The first one comes from the numerical discretisation scheme.
To exclude the influence of statistical variability, we perform simulations with the same $\mathcal{N}(x,t)$ for all cases, which is generated before the simulations by fixing $N_q^t$ (see Appendix\,\ref{App_noise_model}) at each time instant. Figure\,\ref{fig_SLE_covergence}\,(a) shows simulation results with $\varphi=10^{-2}$ at three time instances for varying grid size. 
These interface profiles in Figure\,\ref{fig_SLE_covergence}\,(a) agree well with each other, indicating that the numerical errors indeed
converge.
 We check the average deviation of each interface profile to that with the finest resolution calculated (i.e., $\triangle x = 0.0094$) shown in Figure\,\ref{fig_SLE_covergence}\,(b), where the second-order convergence, as expected, is confirmed, with the prefactor depending, unsurprisingly, on noise strength. 

Since convergence studies for noisy systems are naturally stochastic, the second target quantity to test convergence is ensemble-averaged; here the ensemble consists of 200 independent simulations.
The ensemble-averaged profiles at three-time instances are plotted in Figure\,\ref{fig_SLE_covergence2}\,(a) for varying grid size. 
Here, all profiles are presented relative to the minimum point (i.e., we plot $h$ against $x-x_\mathrm{min}$).
Then we check the average deviation of each ensemble-averaged profile to that with the finest resolution (i.e., $\triangle x = 0.0094$); see Figure\,\ref{fig_SLE_covergence2}\,(b), where the convergence is confirmed. 
Note the deterministic result (black lines) follow second-order convergence, while, as also expected, the stochastic cases converge with approximately first-order convergence.

\section{Numerical verification: thermal capillary waves }\label{sec_ver}
In this section, the numerical scheme is verified by comparison to analytical results known for thermal capillary waves, with particular attention paid to  the effect of the non-uniform grid, which are known to create spurious effects in some stochastic partial differential equation systems \cite{hairer2011approximations}.

The fluctuations of the interface in thermal equilibrium \cite{mecke2005thermal,zhang2021thermal} can be described by the classical theory for thermal capillary waves \cite{aarts2004direct,fetzer2007thermal} with the static spectra of each surface mode (wavenumber) derived from the equipartition theorem.
To capture the development of thermal capillary waves towards the equilibrium state, more advanced theories that include the time evolution of wave spectra were derived by applying linear stability analysis to the SLE \cite{ diez2016metallic, zhang2019molecular}; we will use the model from Ref.\,\cite{zhang2019molecular} to validate our numerical solutions. 
Starting with a smooth initial surface and neglecting disjoining pressure ($A=0$), in dimensionless form the time evolution of the spectrum is given by (see Appendix\,\ref{app_TCW_derive} for derivation):
\begin{equation}
\label{eq_SLE_TCW2} 
|H|_\mathrm{rms} = \sqrt{ - \frac{\varphi L}{ k^2} \left( e^{-2 k^4 t }-1 \right)}\,,
\end{equation}
where $|H|_\mathrm{rms}$ represents the root-mean-square of wave perturbations in the frequency domain, $k$ is the dimensionless wave number and $L$ is the dimensionless length of the film.
\begin{figure}[h]
\centering
\captionsetup{justification=raggedright}
\includegraphics[width=0.68\textwidth]{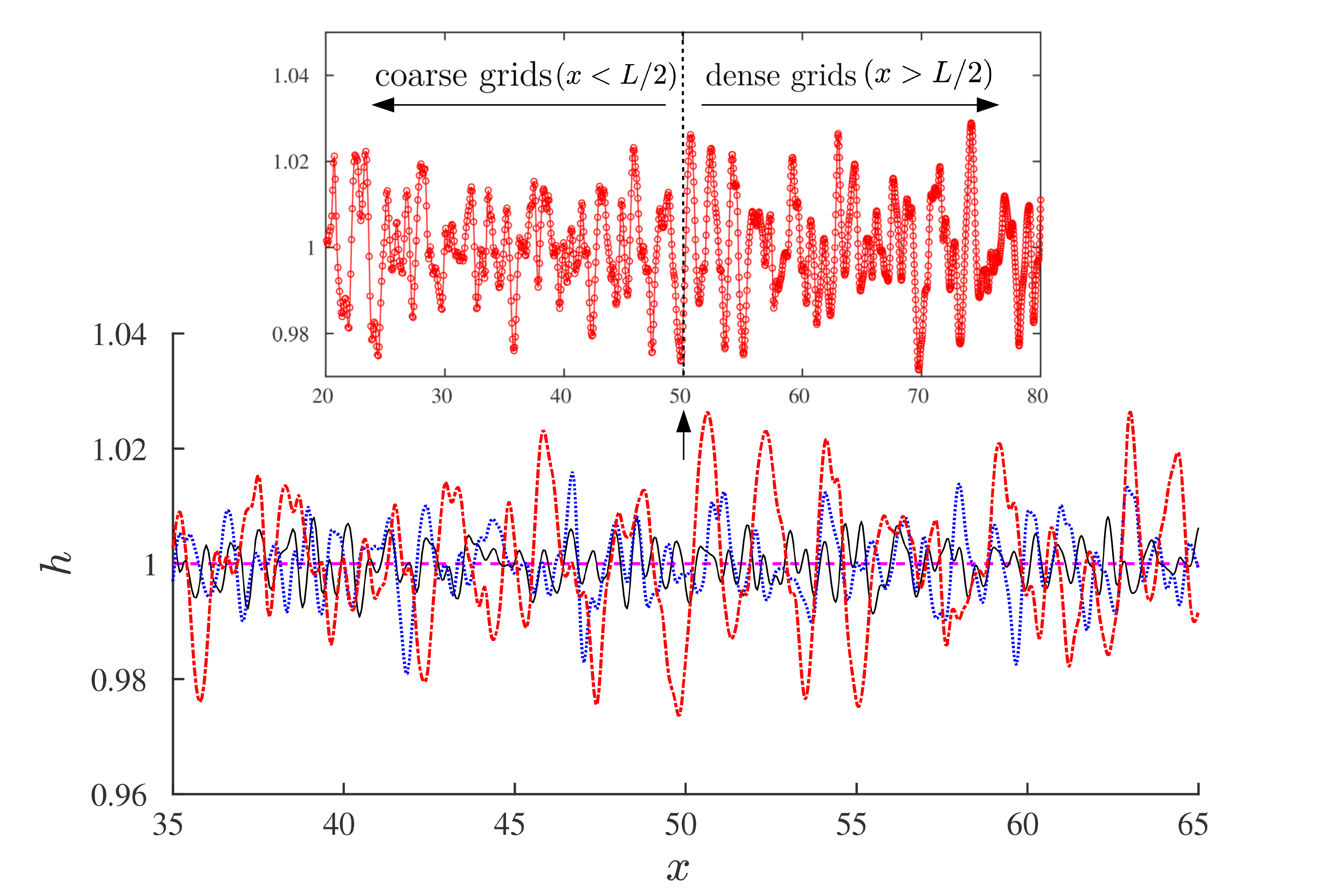}
	\caption{Interface profiles of a bounded film at three time instants, 
	i.e., $t_1=0.0$ (purple dashed line), $t_2=0.001$ (black solid line), $t_3=0.1$ (blue dotted line) and $t_4=1.0$ (red dash-dotted line). The parameters chosen are $\varphi=10^{-3}$ and $L=100$.	
The inset shows how the waves develop on the non-uniform grid (with exponential distribution and markers representing nodal positions), with a higher density as one goes from left to right.}
\label{fig_SLE_TCW1}	
\end{figure}

\begin{figure}[h]
\centering
\captionsetup{justification=raggedright}
\includegraphics[width=0.55\textwidth]{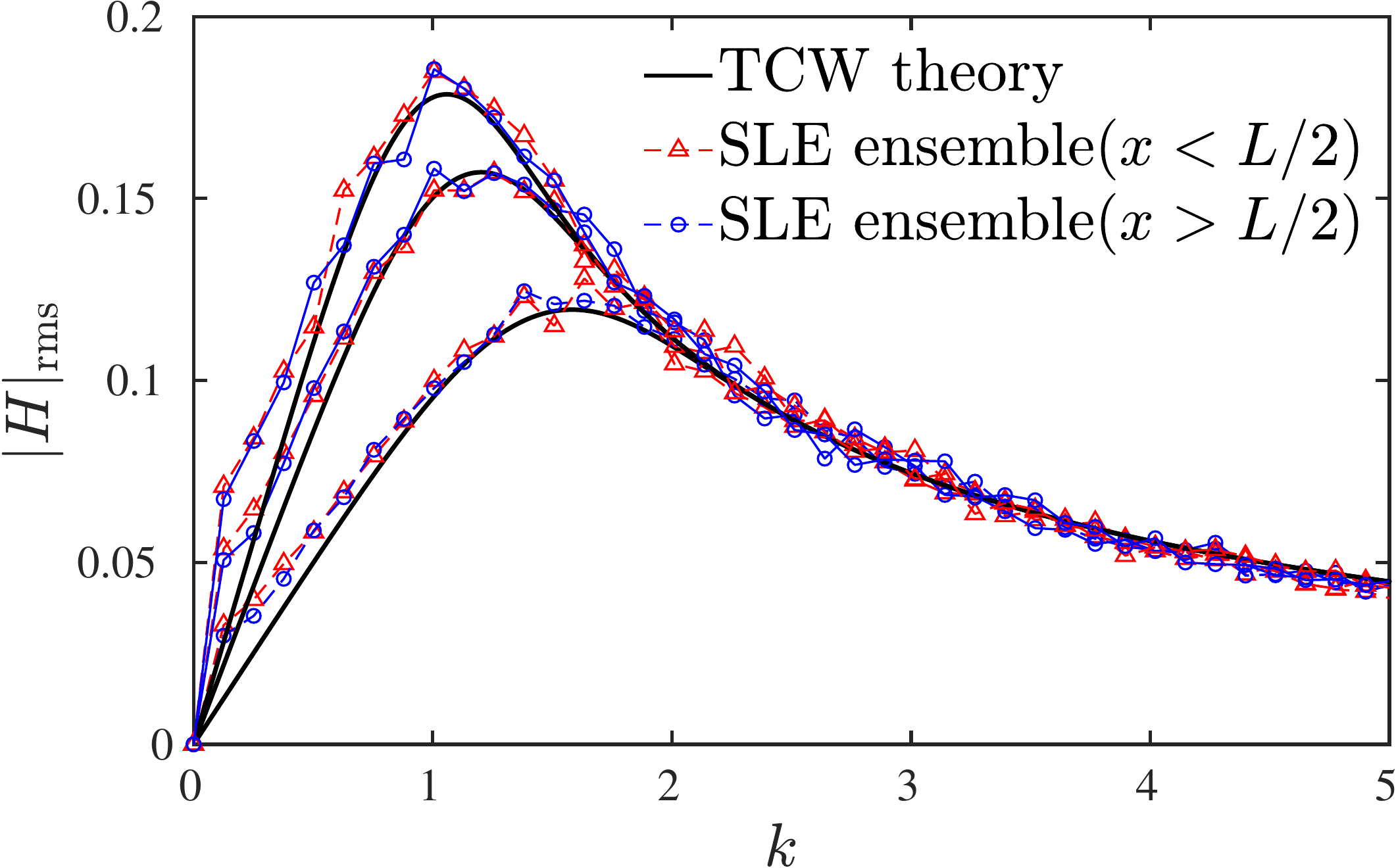}
	\caption{The root-mean-square of (wave) disturbance amplitude versus the wavenumber at three time instants, i.e., $t_1 = 0.2$, $t_2 = 0.4$ and $t_3 = 0.6$; a comparison of ensemble-averaged SLE simulations (dashed lines with markers) and the analytical result (solid lines).}
\label{fig_SLE_TCW2}	
\end{figure}

Figure\,\ref{fig_SLE_TCW1} shows a typical simulation result of the development of thermal capillary waves, where
perturbations, driven by thermal fluctuations, grow against time to generate significant capillary waves at the later stage (see the dash-dotted red line in Figure\,\ref{fig_SLE_TCW1}).
Note that, to test our non-uniform grid implementation, the grid nodes are non-uniformly distributed with the largest grid size $\triangle x_\mathrm{max}=0.1$ at $x=0$ and the smallest grid size $\triangle x_\mathrm{min}=0.001$ at $x=L$.
We use the spatial correlated noise with $L_\mathrm{c}=0.1 \geq \triangle x$. The noise is uncorrelated in time with $\triangle t=10^{-3}$.

To gather statistics, 50 independent simulations (or `realisations') are performed. The solution is divided into two regions: $x<L/2$ with coarse grids and $x>L/2$ with dense ones, to check the influence of the different grid sizes.
On each region, a discrete Fourier transform of $h(x,t)$ is applied to get the power spectral density. We then ensemble average the power spectral density at each time instant over the realisations and take the square root to produce the numerical results in Figure\,\ref{fig_SLE_TCW2} (dash lines with markers).
Good agreement with the theory for thermal capillary waves (Equation\,(\ref{eq_SLE_TCW2})) can be found for both regions (see red and blue line), validating the accuracy of our numerical scheme on extreme non-uniform grids.

\section{Numerical Simulations for Physical Insight }\label{sec_fur}
Having verified our numerical scheme, in this section we use the SLE solver established to explore three bounded thin-film flows with strong nonlinear dynamics: (i) droplet spreading on a precursor film; (ii) sessile droplet coalescence and (iii) thin-film rupture, all of which are ubiquitous phenomena observed both in nature and many industrial processes \cite{herminghaus2008wetting, bonn2009wetting}.

\subsection{Droplet spreading on a precursor film}\label{sec_spreading}

\begin{figure}[h]
\centering
\includegraphics[width=0.7\textwidth]{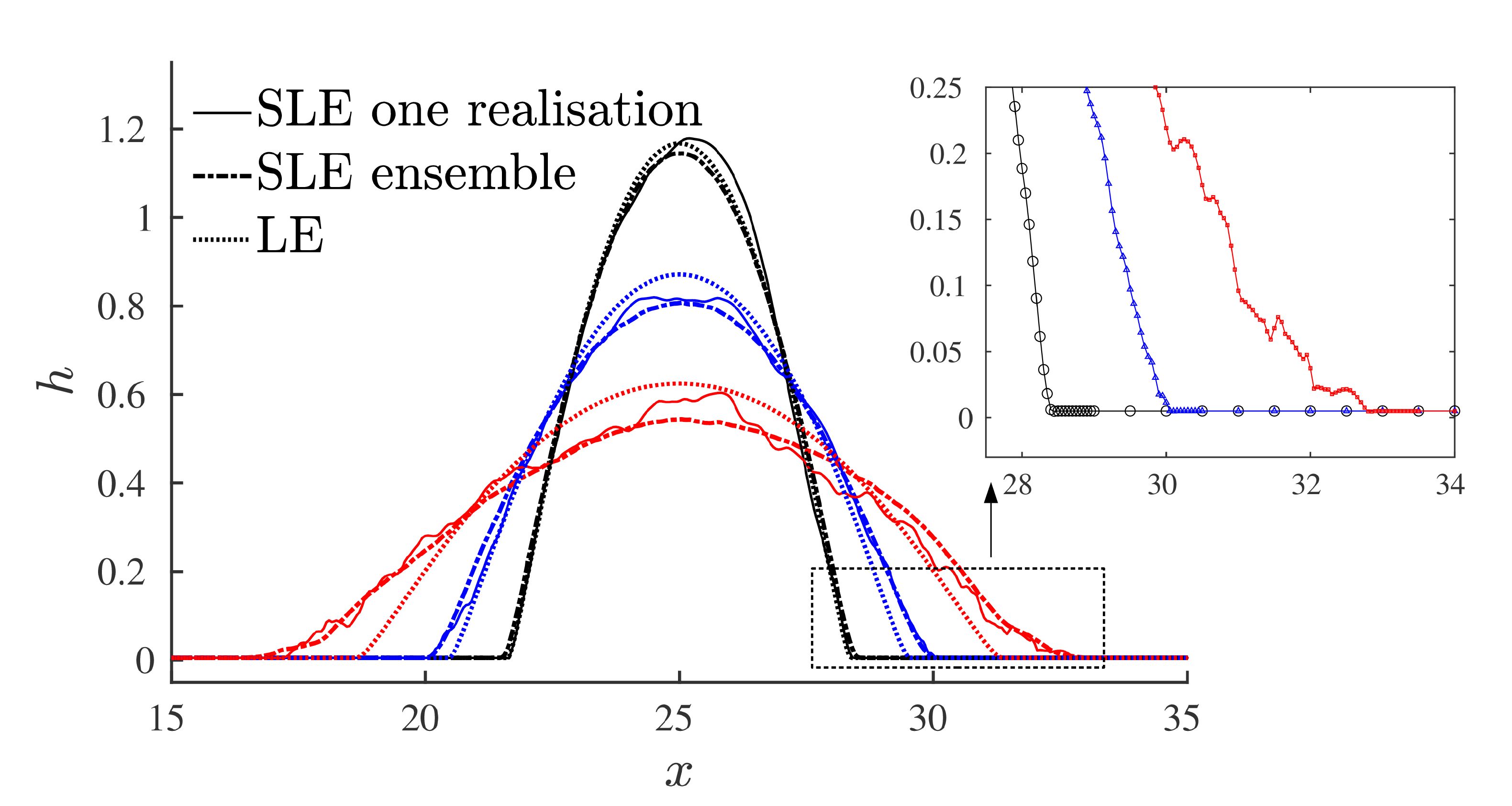}
	\caption[droplet profiles at three time instants with
different boundary conditions.]{Spreading profiles at three time
instants, i.e., $t_1 = 10$ (black lines), $t_2 = 10^2$
(blue lines) and $t_3 = 10^3$ (red lines) with the no-slip
boundary.  where the solid lines represent one
selected realisation. The dash-dotted lines are the average from 50
realisations. The dotted lines are the (LE) solution of the deterministic equation.
The inset shows the local behaviours of the selected realisation (solid lines in the full plot). The lines symbols represent the adaptive grid nodes. }
\label{fig_sp_profile}	
\end{figure}

We now consider the nonlinear dynamics of a droplet spreading on a precursor film, for which the deterministic scaling is given by Tanner's law \cite{tanner1979spreading}, and the more recent work of Davidovitch \cite{davidovitch2005spreading} resulted in a law that accounted for thermal fluctuations.  In particular, these are given by
 \begin{align}
 \label{eq_sp_similarity2}
 & \left\{  
  \begin{array}{ll}
    \ell \sim t\,^{1/7}, \quad \textrm{Hydrodynamic (deterministic) spreading (Tanner's law) \,\,\cite{tanner1979spreading}} , \\
    \ell \sim t\,^{1/4}, \quad \textrm{Spreading driven by thermal fluctuations (Fluctuation-enhanced Tanner's law) \cite{davidovitch2005spreading}}\,\,.  
\end{array}
\right. 
\end{align}
Here, $\ell$ represents a characteristic lateral scale, which is estimated using the average second moment of $h$ (see Equation\,(\ref{eq_sp_width_formula}) in Appendix\,\ref{APP_spreading_law} ) \cite{davidovitch2005spreading}.
The larger power law ($1/4$) in Davidovitch's theory demonstrates that the nanoscale spreading is enhanced by thermal fluctuations. 
However, these two theories are only valid for spreading with the no-slip boundary condition ($l_\mathrm{s} = 0$), which is often inaccurate at the nanoscale.  Therefore, we use the SLE with the slip boundary condition ($l_\mathrm{s} \neq 0$) to derive `slip-modified' power laws for the spreading (see Appendix D for the detailed derivation):
 \begin{align}
 \label{eq_sp_similarity3}
 & \left\{  
  \begin{array}{ll}
    \ell \sim t\,^{1/6}, \quad \textrm{Slip-modified Tanner's law,} \\
    \ell \sim t\,^{1/3}, \quad \textrm{Slip-modified fluctuation-enhanced Tanner's law} .   
\end{array}
\right. 
\end{align}
Note that the new power laws are larger than those based on the no-slip boundary condition, predicting that the spreading is accelerated by slip, as we might expect.

The initial droplet profile for the numerical solutions is given by a section of a sinusoidal function with a precursor film set over the whole simulation domain.
This approach has been widely used \cite{diez2016metallic, duran2019instability, davidovitch2005spreading,nesic2015dynamics} and designed not only for numerical convenience, but also to circumvent the contact-line dynamics (which are fascinating, but not the focus of this article).
To avoid rupture of the thin precursor film (the thickness of precursor film is less than one percent of the initial height of the droplet), we follow Ref.\,\cite{davidovitch2005spreading} and switch off the thermal fluctuations on it. 
Notably, it is unclear whether this `artificial limiter' modifies the numerical path in the stochastic process.
More advanced approaches, like a Brownian bridge technique \cite{russo2021finite} or the inclusion of disjoining pressures, are certainly worthy of future investigation in this regard, but given the slightly artificial nature of the precursor film here they are considered beyond the scope of this article.
When the droplet spreads to a precursor-film node and `pulls it up', namely, $h_i > h^*$, the fluctuations on this node are activated.
Moreover, since we are only interested in the spreading driven by surface tension and fluctuations in this work, the disjoining pressure is neglected ($A=0$).

To accurately capture the spreading power-laws, one needs to simulate the drop spreading for several decades of length. Thus it is computationally expensive to perform simulations on grids uniformly distributed in both the `drop' and `precursor film', in which the height of many nodes will vary very little over each time step. 
To reduce computational costs, we utilise our non-uniform adaptive grids with $\triangle x_\mathrm{min}=0.01$ on the `drop' and $\triangle x_\mathrm{max}=0.1$ on the `film', with the nodes in front of the advancing contact angles refined automatically ($\triangle x_i = \triangle x_\mathrm{min}$) to capture the spreading dynamics (see the inset in Figure\,\ref{fig_sp_profile}). 
Here, each SLE realisation on the uniform grids need about 4 core hours, while one simulation on the non-uniform adaptive grids costs less than 0.2
core hours; i.e., the speed-up is about 20.

\begin{figure}[h]
\centering
\includegraphics[width=0.55\textwidth]{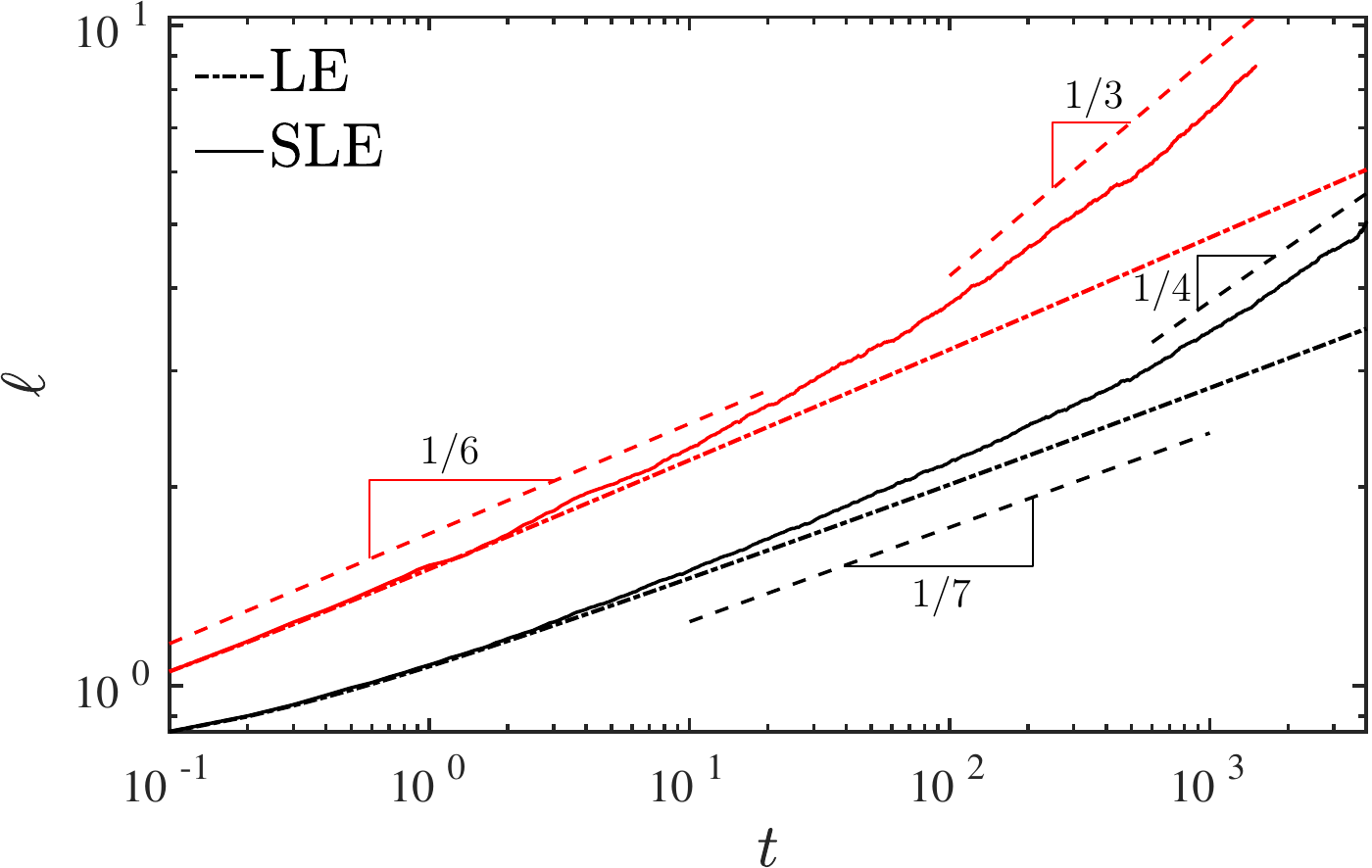}
	\caption{Influence of the slip on the characteristic lateral scale, where black lines and red lines represent no-slip and slip results respectively. The slip length $l_\mathrm{s} = 1.0$. The dash-dotted lines are the solutions of the deterministic cases. The solid lines
are the solutions of the stochastic cases with
$\varphi=10^{-3}$. Dashed lines represent similarity solutions in
Equation\,(\ref{eq_sp_similarity2}) and
(\ref{eq_sp_similarity3}). }
\label{fig_sp_width3}	
\end{figure}

Figure\,\ref{fig_sp_profile} shows the droplet profiles at different
time steps, where the stochastic profiles are the average of the 50
independent realisations with $\varphi=10^{-3}$. 
Note that the initial spreading is much `faster' than that at the later stage (
$t_3-t_2$ is much larger than $t_2-t_1$) due to the initially stronger capillary forces (from the larger curvatures near where the drop meets the precursor), while at the later stage the thermal fluctuations play a significant role and accelerate the process (see the comparison between dotted lines and dash-dotted lines in Figure\,\ref{fig_sp_profile}).  

The characteristic lateral scales (i.e., the drop's width) are plotted in Figure\,\ref{fig_sp_width3}, where the numerical solutions match not only previous analytical solutions with the no-slip boundary condition (black lines) \cite{davidovitch2005spreading}, but also our new `slip-modified' power laws (red lines) very well, giving us further confidence that our numerical scheme captures nanoscale flow physics both accurately and efficiently.
Notably, in both boundary conditions, there exists a transfer from hydrodynamic spreading (Tanner's law) to the fluctuation-driven spreading (Davidovitch's law), showing that the noise dominates over deterministic relaxation only at the later stages of the spreading ($t \gg
1$).  

\subsection{Bounded droplet coalescence}\label{sec_coal}
Previous studies for droplet coalescence on a substrate have been carried out in the thin-film regime \cite{hernandez2012symmetric, ristenpart2006coalescence, karpitschka2014sharp} without any influence of thermal fluctuations taken into account (i.e., using the LE), where surface tension is considered as the main driving force.
However, Perumanath et al \cite{perumanath2019droplet} have shown that the fluctuations are crucial to the dynamics of coalescence of two `free' nano-droplets in a vacuum with MD. A similar influence of noise can be expected in the coalescence of two `bounded' nano-droplets, which will be explored by the SLE solver in this section.
To verify our numerical predictions, we perform independent MD realisations for both symmetric-drop and asymmetric-drop coalescence (see details of MD settings in Appendix\,\ref{APP_MD_setting}).
Note that we focus on 2D cases here.

\begin{figure}[h]
\centering
\includegraphics[width=1.0\textwidth]{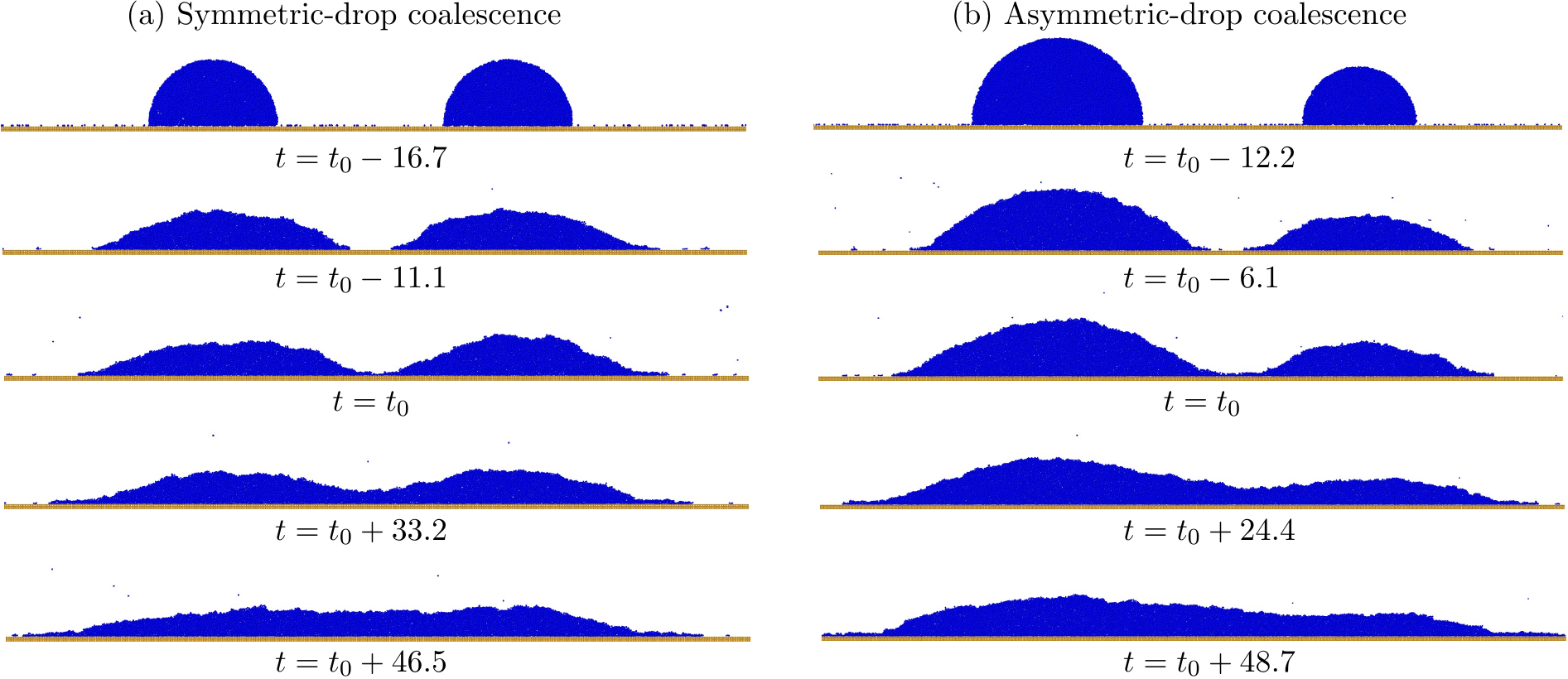}
	\caption{MD results for the coalescence. (a) Symmetric-drop coalescence with the same initial droplet radius (at $t=t_0-16.7$), $R=10\,$nm 
	(b) Asymmetric-drop coalescence with different initial droplet radii (at $t=t_0-12.2$), i.e., $R_1=15\,$nm and $R_2=10\,$nm. }
\label{fig_coal_MD}	
\end{figure}

The MD results of two coalescence cases are presented in Figure\,\ref{fig_coal_MD} with two separate droplets on the substrate
set as the initial conditions.  
The initial distance between the two droplets is $50\,$nm.  Because of the fully wettable substrate, both
droplets spread first until their contact lines meet. The moment when the
two droplets first connect and form a `molecular bridge' is defined as $t_0$, shown in Figure\,\ref{fig_coal_MD}.  After that, the liquid bridge grows and the two droplets eventually merge into one.  

\begin{figure}[h]
\centering
\includegraphics[width=0.7\textwidth]{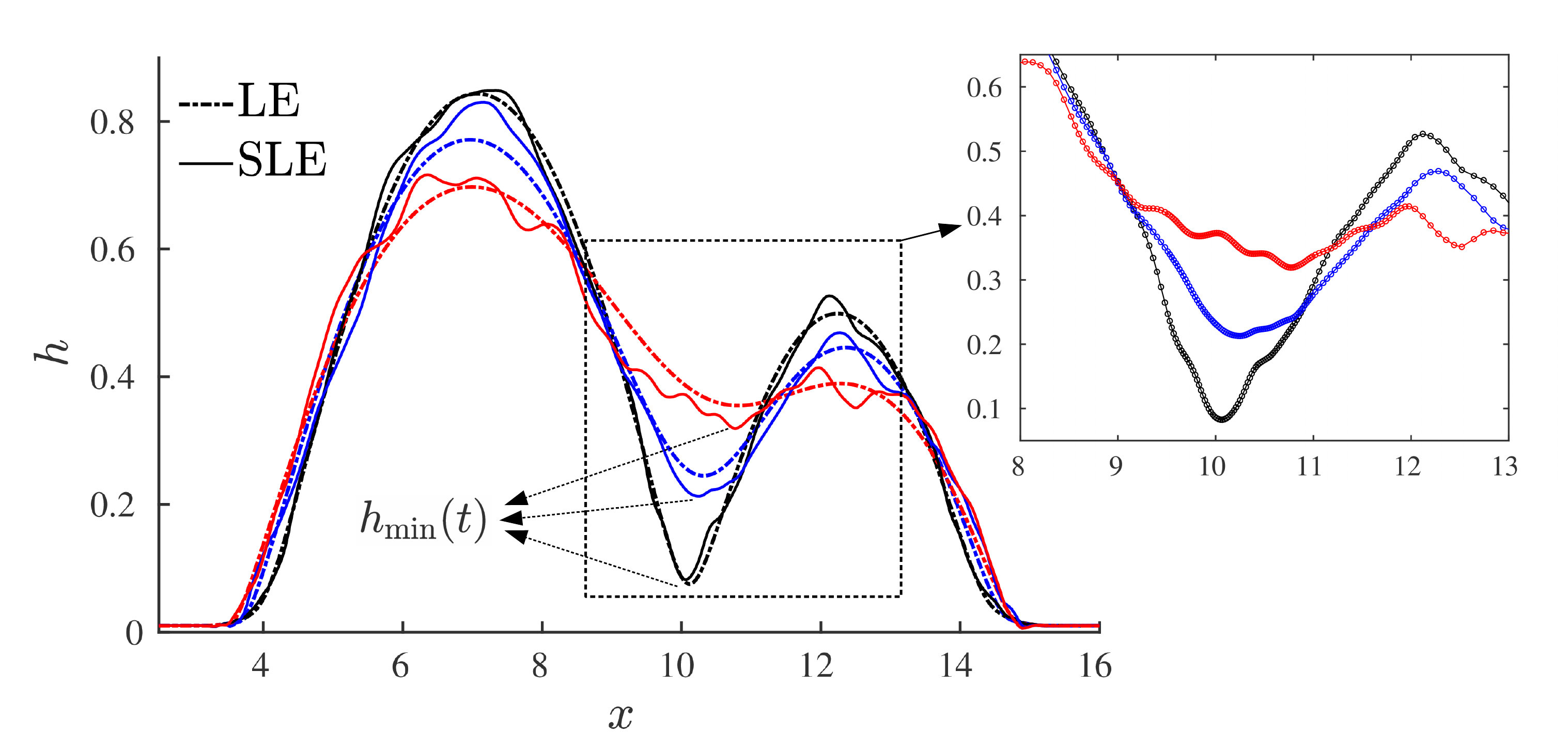}
	\caption{Asymmetric-drop coalescence profiles predicted in a representative simulation by the LE (dash-dotted lines) and SLE (solid lines) at three time instants: $t_1=0.8$, $t_2 = 6.4$ and $t_3 = 13.8$. The inset shows the grid nodes distributions.}
\label{fig_coal_SLE}	
\end{figure}

Similar to Sec.\,\ref{sec_spreading}, here the disjoining pressure is also neglected ($A=0$).
The dimensionless fluctuation intensity ($\varphi = 5.1 \times 10^{-3}$) is calculated from the liquid transport properties of MD (see the details in Appendix\,\ref{APP_MD_setting}).
To set the initial configuration for the SLE, $h(x,0)$, the MD interface profile of each realisation at $t_0$ is extracted from uniform bins along the $x$-axis based on a threshold density. 
Then we shift all the coalescence points to the same position and use the averaged interface profiles as $h(x,0)$ for the SLE.
Simulations are carried out on fixed non-uniform grids with more nodes interpolated near the coalescence point of $h(x,0)$ (see the inset in Figure\,\ref{fig_coal_SLE}). 
The grids are exponentially distributed with $\triangle x_\mathrm{min} = 0.01$ at $h_\mathrm{min}(t_0)$ and $\triangle x_\mathrm{max} = 0.1$ on the precursor film on both sides to capture large gradients near the coalescence point. 
Additionally, we set the spatial correlation length $L_\mathrm{c}=0.1 \geq \triangle x$ and use a constant time step $\triangle t=10^{-3}$. The numerical process is stable with this time step for the configuration of coalescence, so the temporally correlated noise model is not activated here.

\begin{figure}[h]
\centering
\captionsetup{justification=raggedright}
\includegraphics[width=1.0\textwidth]{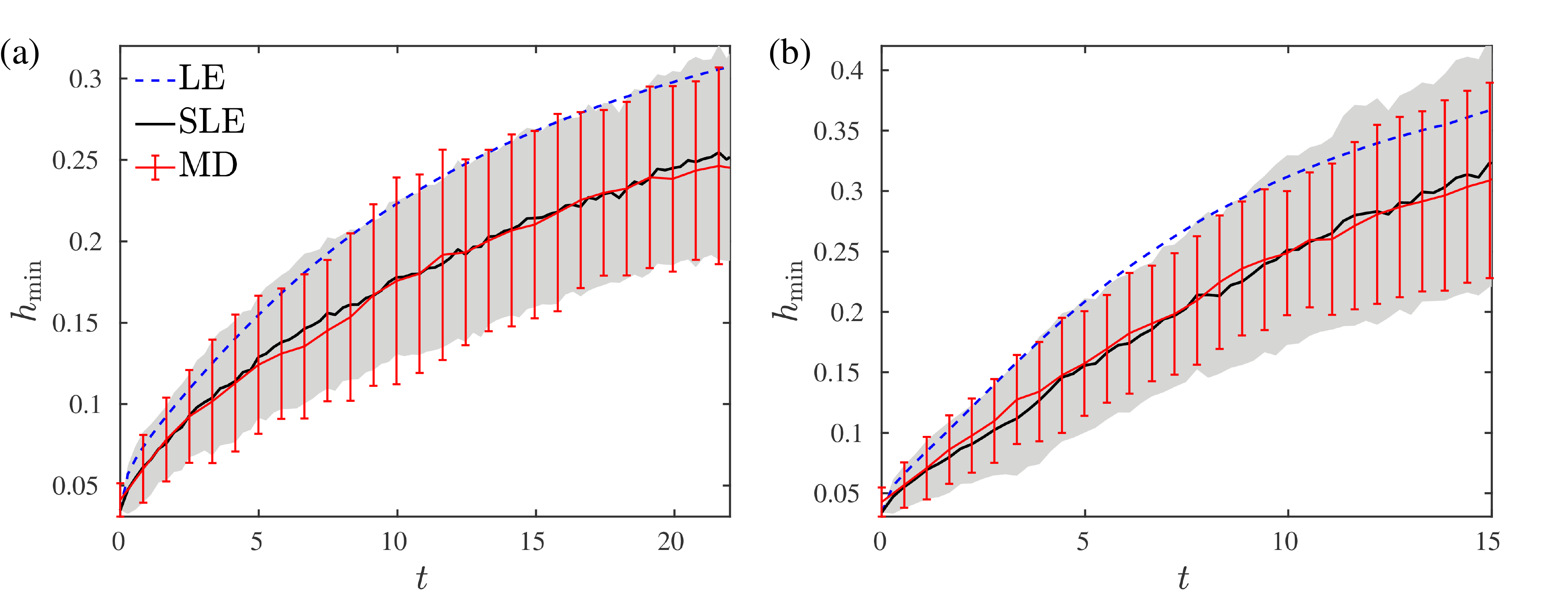}
	\caption{Time evolution of the bridge height during drop coalescence:
comparison between MD and the numerical solutions for the
LE and SLE, for symmetric-drop coalescence in (a) and asymmetric-drop
coalescence in (b). The SLE results are an average from 50
realisations and the MD ones come from 20 realisations. The error bars and shadows in (a) and (b) represent the standard
deviations of the MD and the SLE, respectively.}
\label{fig_coal_hmin}	
\end{figure}
\begin{figure}[h]
\centering
\captionsetup{justification=raggedright}
\includegraphics[width=1.0\textwidth]{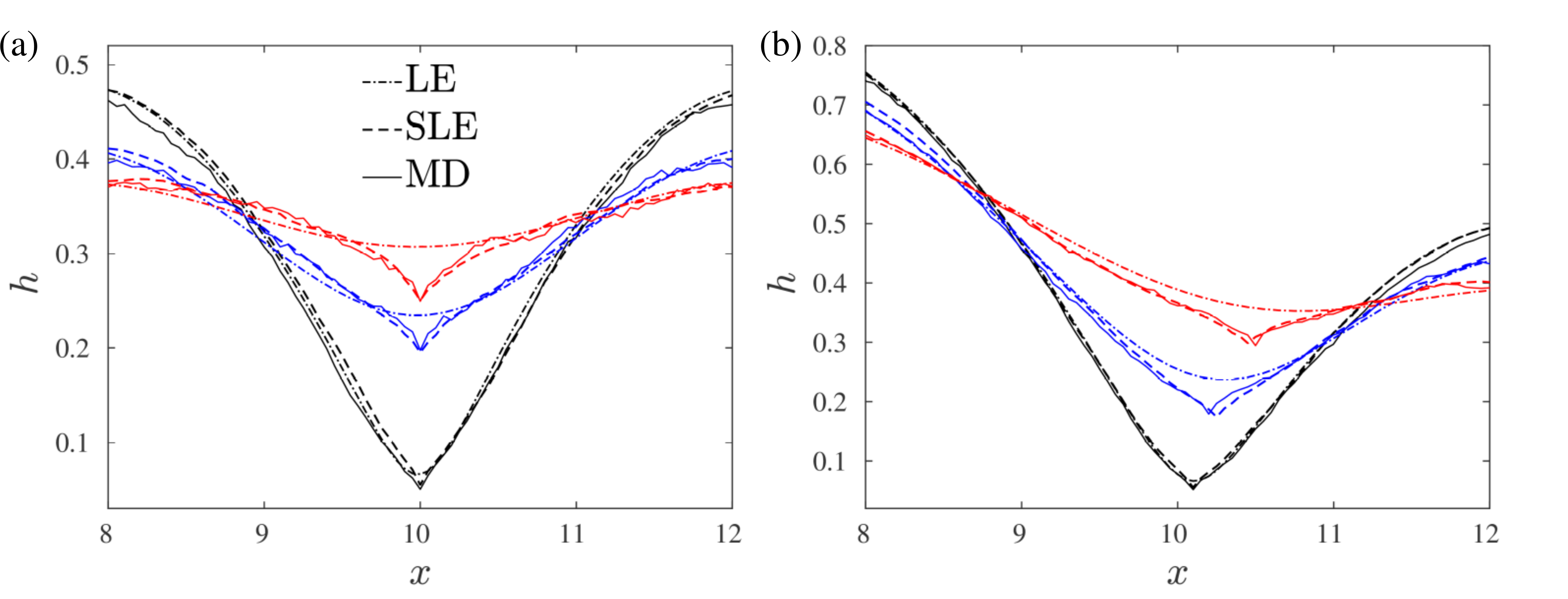}
	\caption{Coalescence profile predicted by the SLE/LE and the MD at three time instants, $t_1$ (black lines), $t_2$ (blue lines) and $t_3$ (red
lines).  (a) symmetric-drop coalescence: $t_1 = t_0+0.83$, $t_2 =
t_0+12.52$ and $t_3 = t_0+22.15$.  (b) Asymmetric-drop
coalescence: $t_1 = t_0+0.83$, $t_2 = t_0+6.37$ and $t_3
= t_0+13.85$.  }
\label{fig_coal_shape}	
\end{figure}

The time evolution of the the minimum bridge height, $h_\mathrm{min}(t)$, is shown in Figure\,\ref{fig_coal_hmin}, where the MD coalescence time, $t_0$ is set as zero for the comparison.  
From a theoretical aspect, $h_\mathrm{min}(t_0)$ is expected to be zero. However in MD, with finite-sized particles, the thickness of the initial `molecular bridge', $h_\mathrm{min}(t_0)$, is approximately equal to the molecular scale. 
In both symmetric and asymmetric cases, good agreement between SLE and MD predictions is found at all times for the mean values, but also, importantly, for the standard deviation. Notably, the deterministic model (LE) is not able to accurately capture the dynamics at these scales, highlighting the significant role that thermal fluctuations must play.
Moreover, the stochastic $h_\mathrm{min}(t)$ always appears smaller than that predicted by the (deterministic) LE for the same time, demonstrating
that noise decelerates the coalescence.  This finding is contrary to previous results for other nanoscale interfacial flows, where typically thermal fluctuations accelerate processes; such as in jet instability \cite{zhao2019revisiting}, thread rupture \cite{zhao2020dynamics} and droplet spreading \cite{davidovitch2005spreading}.
Therefore, it is concluded that the thermal noise cannot always assumed to be a driving force in the interface dynamics at the
nanoscale, and its role is determined by the particular fluid configuration. 
Hern{\'a}ndez-S{\'a}nchez et al. proposed a power law from the LE to describe coalescence dynamics and demonstrated their model with experiments in \cite{hernandez2012symmetric}.
However,  this power law is not found in Figure\,\ref{fig_coal_hmin}, even in the LE solutions (black dashed lines in Figure\,\ref{fig_coal_hmin}). 
At present, the reason is unclear and should be the subject of future investigation.

The ensemble-averaged profiles plotted in Figure\,\ref{fig_coal_shape} show good overall agreement between the MD results (solid lines) and the SLE solutions (dashed lines).
For the asymmetric case, the averaged bottom point ($h_\mathrm{min}$) moves towards the smaller droplet with time, driven by the surface tension.  
As noted above, the deterministic predictions (dash-dotted lines) do not capture the details of the profiles of the MD results, while the SLE results show remarkably good agreement.
Moreover, the SLE solution can reproduce the MD result at a fraction of the computational cost of MD. For the asymmetric-drop coalescence simulation in this section, one MD realization takes more than 200 core hours, while one SLE simulation takes less than 0.2 core hour; i.e., the speed-up is about $10^3$.

\subsection{Thin-film rupture}\label{sec_rup}
The bounded film can become unstable (and rupture) due to the van der Waals forces, as found in experiments with polymeric liquids \cite{becker2003complex}. 
This phenomenon has been successfully described by the LE \cite{becker2003complex} with a similarity solution, $h_\mathrm{min}(t) \sim (t_\mathrm{b}-t)^{1/5}$ \cite{zhang1999similarity}, where the van der Waals forces are modelled by the disjoining pressure term ($\partial \Pi / \partial x$) in Equation\,(\ref{eq_nondim_SLE}). As the scale at the final stage of the rupture reaches several nanometers, thermal fluctuations are expected to play a significant role in rupture with a potentially modified similarity solution due to the fluctuations, which is, remarkably, not included in previous studies \cite{becker2003complex,nesic2015fully,diez2016metallic,duran2019instability}.
Therefore, we use our SLE solver in this section to investigate the influence of thermal fluctuations on the rupture dynamics.  

\begin{figure}[h]
\centering
\captionsetup{justification=raggedright}
\includegraphics[width=0.7\textwidth]{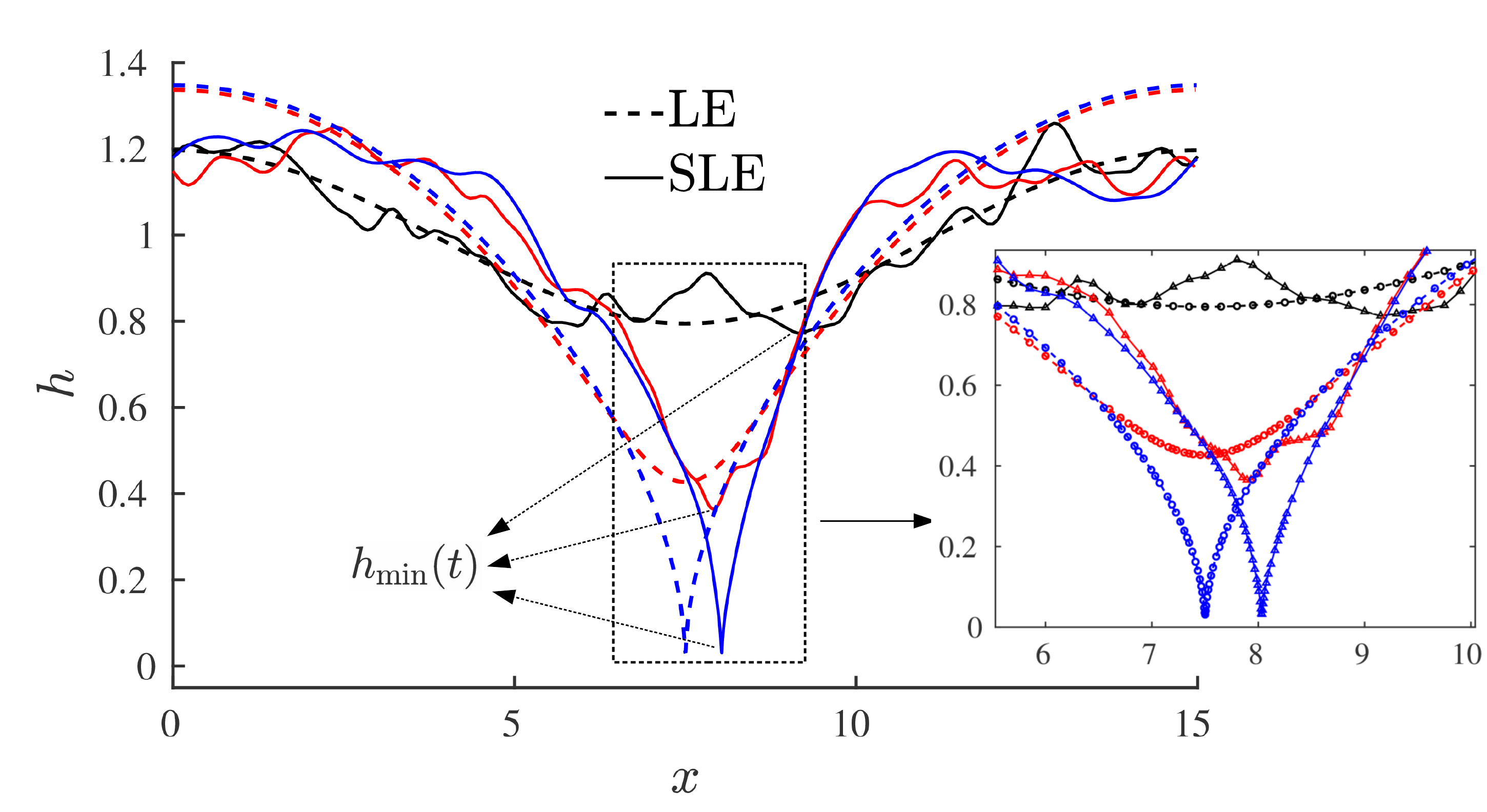}
	\caption{Thin-film rupture profiles at three time instants. The dashed lines are the (deterministic) LE result with $t_1 = 1.0$ (black), $t_2 = 51.0$ (red) and $t_3=52.6$ (blue). The solid lines are from one realisaton for the SLE ($\varphi = 10^{-3}$) with $t_1 = 1.0$ (black), $t_2 = 13.0$ (red) and $t_3=14.0$ (blue). The inset shows the adaptive grids at the rupture point.}
\label{fig_rupture_shape}	
\end{figure}

\begin{figure}[h]
\centering
\captionsetup{justification=raggedright}
\includegraphics[width=0.55\textwidth]{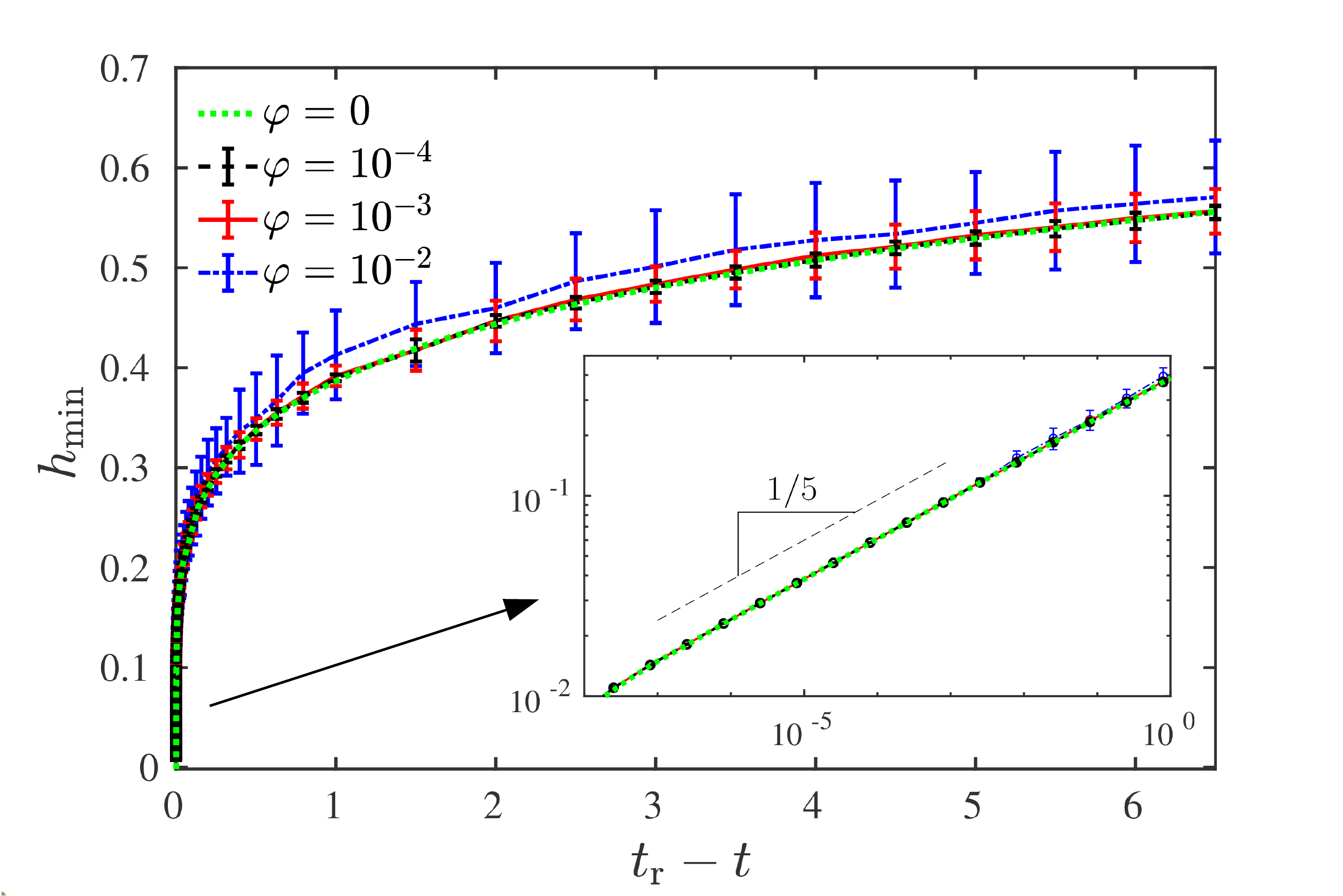}
	\caption{The temporal evolution of the minimum film height for different values of $\varphi$. Here, green lines represent the simulation result of LE. Black, red and blue lines are the average of 50 SLE realisations, where the error bars are the standard deviation. Inset shows the comparison with the similarity solution \cite{zhang1999similarity}.}
\label{fig_rupture_hmin}	
\end{figure}

We set the correlated length $L_\mathrm{c}=0.15$ and correlated time scale $T_\mathrm{c}=10^{-5}$.
The dimensionless Hamaker constant $A=0.2$.
As $\Pi \sim 1/h^3$, the disjoining pressure term increases rapidly at the final stage of the rupture, resulting in two challenges for capturing the dynamics numerically: (i) the spatial derivatives are very large at the rupture point (see the sharp `spikes' in Figure\,\ref{fig_rupture_shape}) and (ii) the breakup happens extremely fast at the final stage (the time for the final breakup, $t_3-t_2$, is much smaller than the time for the early perturbation development, $t_2-t_1$, in Figure\,\ref{fig_rupture_shape}).
To overcome these challenges, we employ both spatial and temporal refinement with the simple criteria $\triangle x \sim h(x,t)$ and $\triangle t \sim h_\mathrm{min}(t)$.
Initially, $\triangle x = 0.15$ and $\triangle t = 10^{-5}$, and they are decreased automatically to $1.5 \times 10^{-4}$ and $10^{-8}$ respectively, to capture the dynamics at the length scale of $10^{-2} h_0$ (see the inset in Figure\,\ref{fig_rupture_hmin}).

Figure\,\ref{fig_rupture_hmin} shows the time evolution of the minimum film height with different fluctuation intensities.
Since we focus on the dynamics near rupture, $h_\mathrm{min}$ is plotted against time to rupture, $t_\mathrm{r}-t$, where $t_\mathrm{r}$ is the time of rupture.
The error bars represent the standard deviation from 50 independent SLE realisations.
It is not surprising to find stronger fluctuations (larger $\varphi$) lead to larger standard deviations, but, it transpires that, these do not affect the `dynamic path' significantly in this regime (see the mean values in Figure\,\ref{fig_rupture_hmin}).
A further interesting finding is that all numerical predictions (with different values of $\varphi$) match the similarity solution proposed by Zhang \& Lister \cite{zhang1999similarity} fairly well to the instant of rupture (see inset in Figure\,\ref{fig_rupture_hmin}), indicating that van der Waals forces, rather than thermal fluctuations, dominate the final stages of thin-film rupture at the nanoscale. This finding is perhaps not surprising given the singular nature of the disjoining pressure term.
The conclusion is also supported by the profiles in Figure\,\ref{fig_rupture_shape}, where the stochastic spike is very similar to the deterministic one, despite the obvious oscillations located on two sides of the film.
Therefore, in this particular regime, thermal fluctuations can play a role in accelerating instability generation, as described in Ref.\,\cite{mecke2005thermal,diez2016metallic,zhang2019molecular}, but once disjoining pressure become significant, it overwhelmes the influence of fluctuations and dominates the dynamics.

\section{Conclusions and future work}
In this work, a simple yet effective numerical scheme for the SLE has been developed to predict the interface dynamics of different classes of nanoscale bounded thin-film flows.
The Fornberg scheme and correlated-noise model is employed in the solver for the non-uniform adaptive grids, offering the capability of capturing local dynamics accurately and efficiently.
Based on verification with theoretical models and comparisons to MD results, this solver is demonstrated to be a powerful tool for studying both linear and nonlinear thin-film flows.

Potential directions for future research are related to the physics predicted by the SLE in spreading, coalescence and rupture. In this article, thermal fluctuations are found to (i) accelerate the droplet spreading, (ii) surprisingly decelerate the bounded droplet coalescence and (iii) not affect the final stages of film rupture. 
In the future, it will be interesting to see whether these findings can be verified experimentally by using colloid-polymer mixture with ultra-low surface tension, which has been applied to generate significant thermal fluctuations in interfacial flows at several micrometers \cite{aarts2004direct,hennequin2006drop, fetzer2007thermal,petit2012break}.
Additionally, the coalescence dynamics  has been shown to be described by the analytical (similarity) solution in the deterministic (LE) cases \cite{hernandez2012symmetric}. So a similar analytical solution is expected to predict the `fluctuation-dominated' coalescence, which would be worthy of further investigations.

Moreover, the results presented in this paper are obtained for two-dimensional flows.
Interesting dynamics in three-dimensional thin-film flows, such as different film-rupture patterns \cite{becker2003complex} and fingering instability in wetting/dewetting \cite{troian1989fingering, baumchen2014influence}, has been thoroughly studied by the three-dimensional LE.
Therefore, it would be interesting to extend our numerical scheme to the three-dimensional SLE and explore the influence of thermal fluctuations on these three-dimensional fluctuating hydrodynamics.
 
\begin{acknowledgments}
This work was supported by the EPSRC (Grants No. EP/N016602/1, No. EP/P020887/1, No. EP/S022848/1 and No. EP/P031684/1).
\end{acknowledgments}

\appendix
\section{The Fornberg scheme}\label{APP_Fornberg}
Here we present a quick derivation of the Fornberg scheme \cite{fornberg1998classroom} to show that the errors in the derivative approximations can be well controlled. Suppose we are given the values of target function $h(x)$ at $x_0<x_1<\ldots<x_n$ and we would like to approximate the $k$th derivative of $h(x)$ at $x\in\Omega$ where $\Omega=[x_0,x_n]$. The Lagrange interpolation polynomial of $h(x)$ based on ${h_i=h(x_i)}_{i=0}^n$, $i=0,1,\ldots,j$ is given by
\begin{equation}
p_n(x) = \sum_{i=0}^{n}\mathcal{L}_{i,n}h_i,
\end{equation}
where $\{\mathcal{L}_{i,n}\}_{i=1}^n$ are the Lagrange polynomials defined by
\begin{equation}
\mathcal{L}_{i,n}(x) = \prod_{j=0,j\neq i}^{n}\frac{x-x_j}{x_i-x_j}.
\end{equation}
It is then immediately known that if $h(x)$ is $n+1$ times continuously differentiable on $\Omega$, then for each $x\in\Omega$,
\begin{equation}
\Vert h^{(k)}(x) - p_n^{(k)}(x)\Vert\leq\Vert \pi^{(k)}(x)\Vert\frac{\Vert h^{(n+1)}(x)\Vert}{k!(n+1-k)!},
\end{equation}
where $k\leq n$, $h^{(k)}$ is the $k$th derivative of $h(x)$, $\pi(x)=(x-x_0)(x-x_1)\cdots(x-x_n)$ and $\Vert$ $\Vert$ denotes the supreme norm on $\Omega$ \cite{howell_derivative_1991}. Thus $h^{(k)}(x)$ can be well approximated by $p_n^{(k)}(x)$ for $k\leq n$ if $\vert x_n-x_0\vert$ is small,
\begin{equation}
h^{(k)}(x)\approx p_n^{(k)}(x) = \sum_{i=0}^{n}\frac{d^k\mathcal{L}_{i,n}(x)}{dx^k} h_i = \sum_{i=0}^{n} \mathcal{L}_{i,n}^k(x) h_i,
\end{equation} 
where $\mathcal{L}_{i,n}^k(x)$ is the expression for the Fornberg coefficients and can be calculated recursively. We know
\begin{equation}
\mathcal{L}_{i,n}(x) = \frac{x-x_n}{x_i-x_n}\mathcal{L}_{i,n-1}(x),\quad\text{for}\quad i\neq n,
\end{equation}
and
\begin{equation}
\mathcal{L}_{n,n}(x) = \frac{\prod_{j=0}^{n-2}(x_{n-1}-x_j)}{\prod_{j=0}^{n-1}(x_n-x_j)}(x-x_{n-1})\mathcal{L}_{n-1,n-1}(x).
\end{equation}
Then for $i\neq n$
\begin{align}
\mathcal{L}_{i,n}^k(x) &= \frac{d^k}{dx^k}\left(\frac{x-x_n}{x_i-x_n}\mathcal{L}_{i,n-1}(x)\right) \\
&= \frac{x-x_n}{x_i-x_n}\mathcal{L}_{i,n-1}^k(x) + \frac{k}{x_i-x_n}\mathcal{L}^{k-1}_{i,n-1}(x),
\end{align}
and
\begin{align}
\mathcal{L}_{n,n}^k(x) &= \frac{d^k}{dx^k}\left(\frac{\prod_{j=0}^{n-2}(x_{n-1}-x_j)}{\prod_{j=0}^{n-1}(x_n-x_j)}(x-x_{n-1})\mathcal{L}_{n-1,n-1}(x)\right) \\
&= \frac{\prod_{j=0}^{n-2}(x_{n-1}-x_j)}{\prod_{j=0}^{n-1}(x_n-x_j)}\left[(x-x_{n-1})\mathcal{L}_{n-1,n-1}^k(x) + k\mathcal{L}_{n-1,n-1}^{k-1}(x)\right].
\end{align}
Since $\mathcal{L}_{0,0}$ represents a single point, it is natural to set $\mathcal{L}_{0,0}^0=0$. Further more $\mathcal{L}_{j,m}^s = 0$ for $s>m$ because the $s$th derivative of a $m$th order polynomial is always zero. Without loss of generality we set $x=0$ to get
\begin{align}
\mathcal{L}_{i,n}^{k} &= \frac{1}{x_n-x_i}\left(x_n\mathcal{L}_{i,n-1}^k-k\mathcal{L}^{k-1}_{i,n-1}\right), \\
\mathcal{L}_{n,n}^k &= \frac{\prod_{j=0}^{n-2}(x_j-x_{n-1})}{\prod_{j=0}^{n-1}(x_j-x_n)}\left[x_{n-1}\mathcal{L}_{n-1,n-1}^k(x) - k\mathcal{L}_{n-1,n-1}^{k-1}(x)\right],\\
\Vert h^{(k)} &- p_n^{(k)}\Vert  \leq\Vert\pi^{(k)}\Vert\frac{\Vert h^{(n+1)}\Vert}{k!(n+1-k)!}\simeq\mathcal{O}({\delta x}^{n+1-k}),
\end{align}
where $\delta x = \max\{\vert x_0 \vert,\vert x_n \vert\}$. Generally speaking the Fornberg scheme has at least accuracy of order $n+1-k$.

\section{Correlated noise model}\label{App_noise_model}
In this appendix, we introduce the spatially correlated noise model, proposed by Gr{\"u}n in \cite{grun2006thin},
where an exponential correlation function is employed
 \begin{align}
F_\mathrm{cor}(x,L_\mathrm{c})=
 & \left\{  
  \begin{array}{ll}
    X^{-1} \mathrm{exp}\left[ -\frac{1}{2} \left( \frac{L}{L_\mathrm{c}} \mathrm{sin}\left( \pi x/L\right)\right)^2 \right], \quad & \mathrm{for}\,\,L_\mathrm{c}>0 \,, \\
    \delta(x) , \quad & \mathrm{for}\,\,L_\mathrm{c}=0 \,.     
\end{array}
\right. 
\label{eq_app_corre}
\end{align}
\begin{figure}[h]
\center
\includegraphics[width=0.55\textwidth]{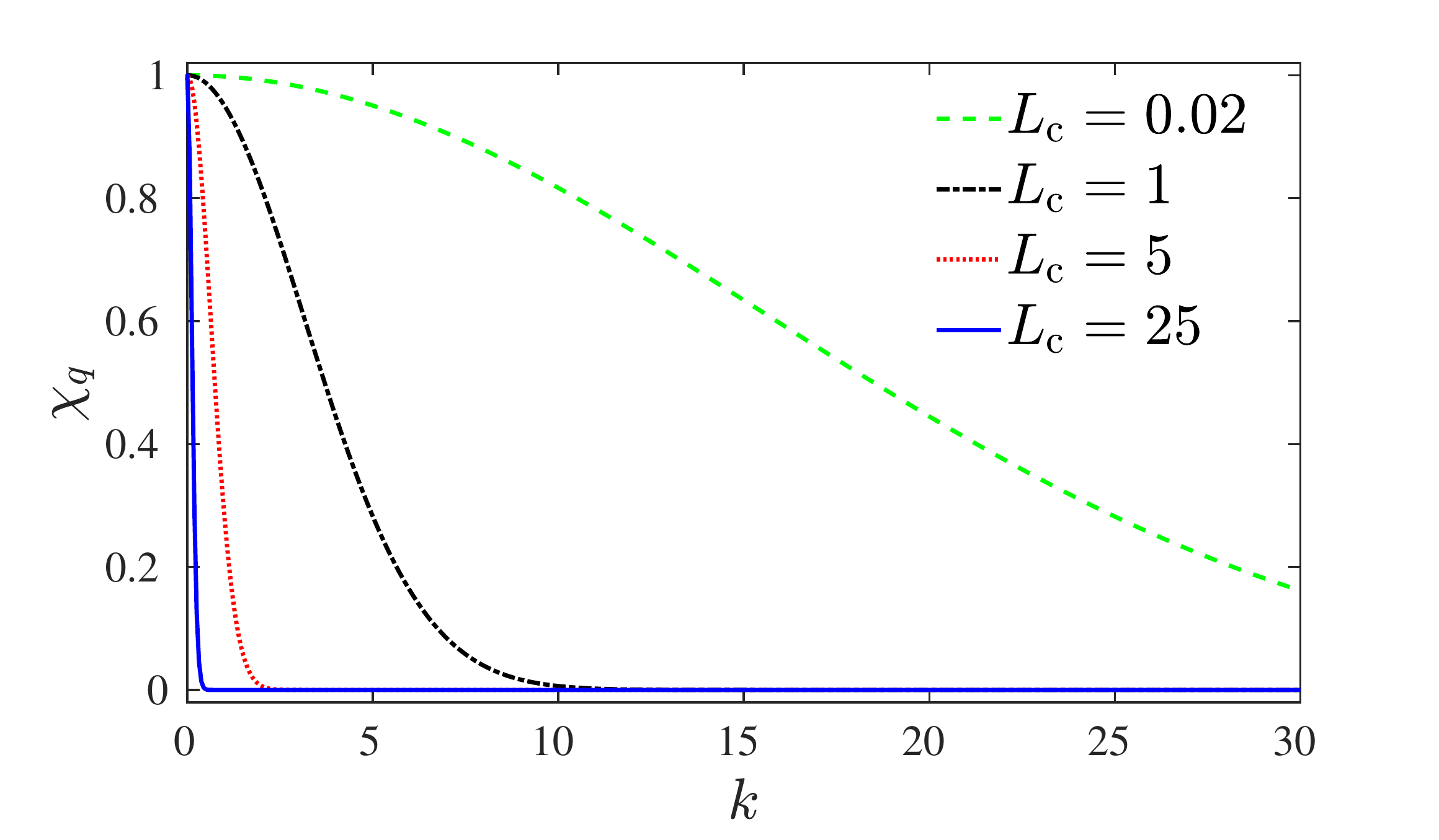} 
    \caption{Linear spectrum of eigenvalues for several values of $L_\mathrm{c}$ from Equation\,(\ref{eq_chi}). Here, the wavenumber, $k=2 \pi q/L$}
\label{fig_chi}
\end{figure}
Here, $L_\mathrm{c}$ is the spatial correlation length, $L$ is the domain length, $X$ is such that $\int_0^L F_\mathrm{cor}(x,L_\mathrm{c}) dx = 1$.

Diez et al.\cite{diez2016metallic} calculated the integral and found that $\chi_q$ could be expressed by the Bessel function,
\begin{equation}
\chi_q = I_k(\alpha) / I_0(\alpha)\,,
\label{eq_chi}
\end{equation}
where
$$\alpha = \left( \frac{L}{2 L_\mathrm{c}}\right)^2\,\quad \mathrm{and} \quad k = 2 \pi q/L \,.$$
Figure\,\ref{fig_chi} shows the eigenvalue spectrum for several values of $L_\mathrm{c}$.
Note that for $L_\mathrm{c} \rightarrow 0$ (i.e., $\alpha \rightarrow \infty$), we have $\chi_q \rightarrow 1$ for all $q$, leading to the limiting case of the white (uncorrelated) noise.

\begin{figure}[h]
\center
\includegraphics[width=0.55\textwidth]{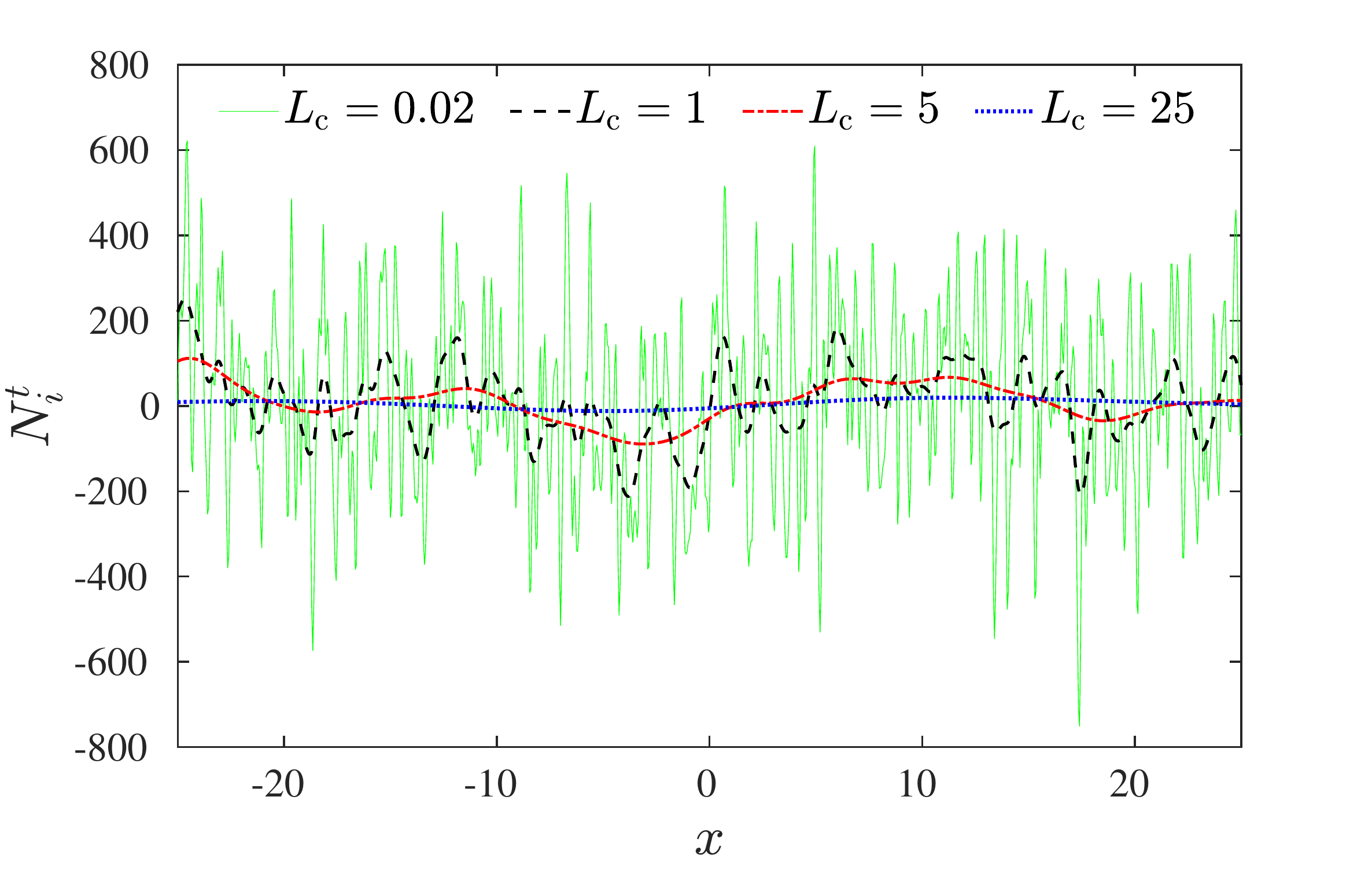} 
    \caption{Spatially correlated noise with different $L_\mathrm{c}$.}
\label{fig_app_noise}
\end{figure}

\begin{figure}[h]
\center
\includegraphics[width=0.55\textwidth]{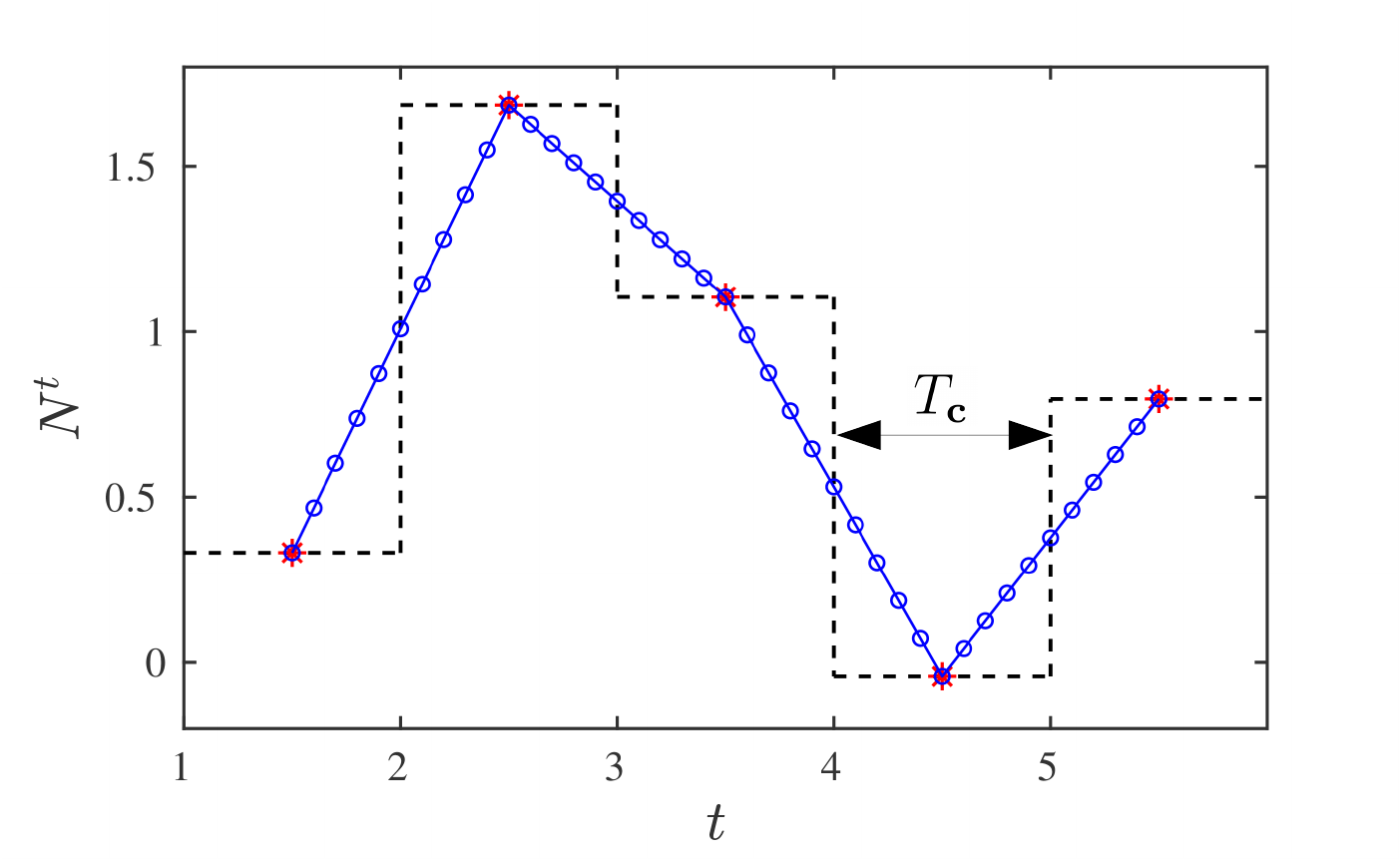} 
    \caption{The temporally correlated stochastic term $N^t$ using linear interpolation. }
\label{fig_app_linear}
\end{figure}

The term $g_q$ corresponds to the set of orthonormal eigenfunctions according to
 \begin{align}
g_q(x)=
 & \left\{  
  \begin{array}{ll}
    \sqrt{\frac{2}{L}} \mathrm{cos}(\frac{2 \pi q x}{L}) , \quad & \mathrm{for}\,\,q>0  \\
    \sqrt{\frac{1}{L}}  , \quad & \mathrm{for}\,\,q=0  \\  
    \sqrt{\frac{2}{L}} \mathrm{sin}(\frac{2 \pi q x}{L}) , \quad & \mathrm{for}\,\,q<0      
\end{array}
\right. 
\label{eq_fourier_domain_app2}
\end{align}
Therefore, the discretised expression of the noise term is 
\begin{equation}
N^t_i= \frac{1}{\sqrt{\triangle t}}\sum^{q = \frac{M+1}{2}}_{q=-\frac{M+1}{2}} \chi_q\, N^t_q\, g_q(x) \,,
\end{equation}
where $M$ is the number of nodes.
Samples of $N^t_i$ are illustrated in Figure\,\ref{fig_app_noise} with different spatial correlation lengths.
Note that a larger $L_\mathrm{c}$ leads to smooth large-wavelength and small-amplitude noise.

The temporally correlated noise model proposed by Zhao et al. \cite{zhao2020dynamics} is employed in this work, shown in Figure\,\ref{fig_app_linear}, where $T_\mathrm{c}$ is the correlation timescale.
When the adaptive time steps are reduced to less than $T_\mathrm{c}$ to capture local dynamics (e.g., for the final stage of the film rupture), 
this temporally correlated model is activated with time step added using linear interpolation.
The discretised expression of the noise term becomes
\begin{equation}
N^t_i= \frac{1}{\sqrt{T_\mathrm{c}}}\sum^{q = \frac{M+1}{2}}_{q=-\frac{M+1}{2}} \chi_q\, N^t_q\, g_q(x) \,.
\end{equation}

\section{Derivation of theory for thermal capillary waves \label{app_TCW_derive}}
Without the disjoining pressure term, Equation\,(\ref{eq_nondim_SLE}) becomes,
\begin{equation}
\label{eq_nondim_SLE2}
\partial_t h= - \partial_x \left( h^3 \partial_x^3 h - \sqrt{2 \varphi h^3}\,  \mathcal{N} \right) \,.
\end{equation}
For the linear instability, we set $h = 1+ \hat{h} $ with $\hat{h}\ll 1$ to linerize Equation\,(\ref{eq_nondim_SLE2}):
\begin{equation}
\partial_t \hat{h} + \partial_x^4 \hat{h}^4 = \sqrt{2 \varphi} \,\partial_x \mathcal{N }.
 \label{ceq_LSLE}
\end{equation}
Then a finite Fourier transform is applied to Equation\,(\ref{ceq_LSLE}) to get 
\begin{equation}
\partial_t H + k^4 H = ik \sqrt{2 \varphi} N \,,
 \label{ceq_FSLE}
\end{equation}
where the transformed variables are defined as follows:
$$H(k,t) = \int^L_0 \hat{h}(x,t) e^{-ikx} dx \quad \textrm{and} \quad N(k,t) = \int^L_0 \mathcal{N}(x,t) e^{-ikx} dx.$$

The solution of  Equation\,(\ref{ceq_FSLE}) is linearly decomposed into two parts:
\begin{equation}
H={H}_\mathrm{LE}+{H}_\mathrm{fluc} \, .
\label{ceq_f2}
\end{equation}
The first part is the solution to the homogenous form of Equation\,(\ref{ceq_FSLE}) (i.e., with $\varphi=0$) with some stationary initial disturbances (i.e., $H$=$H_\mathrm{i}$ at $t=0$). Since we start from a smooth initial surface,  $H_\mathrm{LE}(k,t)=0$.
The second component of the solution arises from solving the full form of Equation\,(\ref{ceq_f2}) without any initial disturbances; this part of the solution is solely due to fluctuations, and is thus denoted $H_\mathrm{fluc}$.
This is obtained by determining the homogeneous equation's impulse response,
\begin{align}
H_\mathrm{res}(k,t) = e^{-k^4 t}\,, 
\label{ceq_f4}
\end{align}
which due to the linear, time-invariant nature of the system, allows to write
\begin{align}
\overline{|H_\mathrm{fluc}|^2} & = \left( ik \sqrt{2 \varphi} \right)^2 
\overline{ \left|
\int^t_0 N(k,t-\tau) H_\mathrm{res}(k,\tau) d \tau
 \right|^2} 
\label{ceq_f5}
\end{align}

$H$ is both a random and complex variable with zero mean.
So the root mean square (rms) of $H$ is sought, which from Equation\,(\ref{ceq_f2}) is given by
\begin{align}
|H|_\mathrm{rms}& =\sqrt{\overline{|H_\mathrm{LE}+H_\mathrm{fluc}|^2}} = \sqrt{\overline{|H_\mathrm{fluc}|^2}} \,.
\label{ceq_f6} 
\end{align}
Because $N$ is uncorrelated Gaussian white noise, and the variance of the norm of the white noise $\overline{|N|^2}=L$, Equation\,(\ref{ceq_f4}) and (\ref{ceq_f5}) combine to give
\begin{align}
\overline{|H_\mathrm{fluc}|^2}
& = 2 \varphi k^2 \int^t_0 \overline{|N(k,t-\tau)|^2} H(k,\tau)^2 d \tau \, , \nonumber\\ 
&=2 \varphi k^2 L \int^t_0  H^2 d \tau \nonumber \,, \\
& =  - \frac{\varphi L}{ k^2} \left( e^{-2 k^4 t }-1 \right)\,.
\label{ceq_f8}
\end{align}
Equation\,(\ref{ceq_f6}) and (\ref{ceq_f8}) constitute the theory for thermal capillary waves used in Sec.\,\ref{sec_ver}.

\section{Derivation for the spreading power-laws}\label{APP_spreading_law}
In this appendix, we present the derivations for both no-slip and slip theories, i.e., $\ell \sim t^n$.
The average second moment of $h$ is applied to estimate the drop's width $\ell$ \cite{davidovitch2005spreading} whose explicit expression is
\begin{equation}
\label{eq_sp_width_formula}
\ell(t) = \left\langle \left[ \frac{1}{V} \int (x-X)^2 h(x,t)  d x \right] \right\rangle\,,
\end{equation}
where $V = \int h dx$ is the constant volume
of the droplet, $X = \left( \int x h
dx\right)/ V$ is the instantaneous position of the
droplet center, and $\left\langle \hdots \right\rangle$ represents the
ensemble average of all the realisations.

In order to decide the power law ($n$), a similarity transform is made with the change of variables:
\begin{equation}
\label{eq_trans_var}
x = b \breve{x}, \qquad h=b^{\alpha} \breve{h}, \qquad t = b^{\eta} \breve{t},
\end{equation}
where $b$ is an arbitrary factor, and $\alpha$ and $\eta$ are
constants that remain to be fixed.  The symbol `$\,\breve{•}\,$' means
`transformed' variables.  According to the scaling relation above ($\breve{x} \sim \breve{h} \sim \breve{t} \sim \mathrm{O}(1)$), we
can easily obtain $\ell \sim x \sim t^{1/\eta}$,
namely, the power law is equal to $ 1/\eta$.  For the value of $\eta$, we substitute the transform relations above into the SLE (Equation\,(\ref{eq_nondim_SLE})) and obtain
\begin{equation}
 \frac{\partial \breve{h}}{\partial \breve{t}} = 
-\left( b^{3 \alpha+\eta-4} \right) \partial_{\breve{x}} \left( \breve{h}^3 \partial^3_{\breve{x}} \breve{h} \right)
+\left[ b^{(\alpha+\eta-3)/2} \right] \sqrt{2 \varphi} \partial_{\breve{x}} \left[  \breve{h}^{3/2} \breve{\mathcal{N}}(\breve{x},\breve{t})  \right]\,.
\label{eq_TF3}
\end{equation}
Here, we have two independent force terms on the right-hand side: (i) the deterministic term due to the surface tension and (ii) the stochastic term due to the thermal fluctuations.
To hold the `similarity' of the transform, the scaling powers of the arbitrary scaling factor $b$ should always be zero, namely,
 \begin{align}
 \label{eq_sp_similarity1}
 & \left\{  
  \begin{array}{ll}
     3 \alpha+\eta-4 = 0 \,,\quad \textrm{for the deterministic term,} \\ 
     \alpha+\eta-3 = 0\,, \quad \textrm{for the stochastic term.}    
\end{array}
\right. 
\end{align}
In addition, no matter which force drives the spreading, the droplet volume,
$$ V = \int h dx =b^{\alpha+1} \int \breve{h} d \breve{x}\,,$$ 
should always be conserved, requiring $\alpha=-1$.  So, we can obtain
the value of the left coefficient $\eta$ from
Equation\,(\ref{eq_sp_similarity1}), i.e., $\eta=7$ in the surface
tension term; and $\eta=4$ in the stochastic term, implying two
power-law spreading regimes:
 \begin{align}
 & \left\{  
  \begin{array}{ll}
    \ell \sim t\,^{1/7}, \quad \textrm{Tanner's law,} \\
    \ell \sim t\,^{1/4}, \quad \textrm{fluctuation enhanced Tanner's law},  
\end{array}
\right. 
\end{align}
which have been proposed by Tanner \cite{tanner1979spreading} and Davidovitch et al. \cite{davidovitch2005spreading}, respectively.

To take the slip effect into account, Equation\,(\ref{eq_sp_similarity1}) is modified:
\begin{equation}
 \frac{\partial \breve{h}}{\partial \breve{t}} = 
-\left( b^{2 \alpha+\eta-4} \right) \partial_{\breve{x}} \left(3 \tilde{\ell_\mathrm{s}} \breve{h}^2 \partial^3_{\breve{x}} \breve{h} \right)
+ b^{(\eta-3)/2} \sqrt{2 \varphi} \partial_{\breve{x}} \left[  \sqrt{3 \tilde{\ell_\mathrm{s}} \breve{h}^2} \breve{\mathcal{N}}(\breve{x},\breve{t})  \right].
\label{eq_TF3}
\end{equation}
By the same approach, we get
 \begin{align}
 \label{eq_sp_similarity4}
 & \left\{  
  \begin{array}{ll}
     2 \alpha+\eta-4 = 0 \,,\quad \textrm{for the deterministic term,} \\ 
     \eta-3 = 0\,, \quad \textrm{for the stochastic term,}    
\end{array}
\right. 
\end{align}
where $\alpha$ is still equal to $-1$. Therefore, we can obtain a `slip-modified' power law for the spreading:
 \begin{align}
 & \left\{  
  \begin{array}{ll}
    \ell \sim t\,^{1/6}, \quad \textrm{Slip-modified Tanner's law,} \\
    \ell \sim t\,^{1/3}, \quad \textrm{Slip-modified fluctuation enhanced Tanner's law}.   
\end{array}
\right. 
\end{align}

\section{MD settings for the droplet coalescence}\label{APP_MD_setting}
In this work, we choose the mW model \cite{molinero2008water} to simulate liquid water.
The model mimics the hydrogen-bonded structure of water through the introduction of a non-bond angular dependent term that encourages
tetrahedral configurations. The model contains two terms: (i) $\phi_{ij}$ depending on the distances between pairs of atoms (represented by $r_{ij}$ and $s_{ik}$) and (ii) $\phi_{ijk}$ depending on the angles formed by triplets of atoms (represented by $\theta_{ijk}$).
The full expression is given by
\begin{align}
\mathcal{U} &= \sum \limits_{i} \sum \limits_{j>i} \phi_{ij}(r_{ij}) 
+ \sum \limits_{i} \sum \limits_{j \neq i} \sum \limits_{k>j} \phi_{ijk}(r_{ij}, s_{ik}, \theta_{ijk})\,, \label{eq_mW} \\ 
& \phi_{ij}(r_{ij})  = A \epsilon \left[ B \left( \frac{\sigma}{r_{ij}} \right)^p - \left( \frac{\sigma}{r_{ij}} \right)^q \right] 
\mathrm{exp}\left(\frac{\sigma}{r_{ij}-a \sigma}\right)\,, \nonumber \\
& \phi_{ijk}(r_{ij}, s_{ik}, \theta_{ijk}) = \kappa \epsilon (\cos \theta_{ijk} -\cos \theta_0)^2 \,
\mathrm{exp} \left( \frac{\chi \sigma}{r_{ij}-a \sigma} \right) \,
\mathrm{exp}\left( \frac{\chi \sigma}{s_{ik}-a \sigma} \right)\,, \nonumber
\end{align}
where $A$, $B$, $p$, $q$, $\chi$ and $\kappa$ respectively give the form and scale
to the potential, and $\theta_0$ represents the tetrahedral angles.
All the parameters are presented in Table\,\ref{tab_mw_par}.
\begin{table}[h]
\caption{\label{tab_mw_par} Parameters of the mW model}
\begin{ruledtabular}
\centering
\begin{tabular}{cccccccccc}
 $\epsilon$\,(kJ\,mol$^{-1}$) & $\sigma$\,(nm) & $A$  & $B$ & $p$  & $q$  & $\chi$ &  $\kappa$ & $a$ & $\theta_0$\,(degree)   \\
\hline
25.87  & 0.2390 &  7.050  & 0.6022 & 4  & 0  & 1.2 & 23.15 & 1.8 & $109.47$   \\
\end{tabular}
\end{ruledtabular}
\end{table}

The platinum substrate is assumed to be rigid with an atomic mass of $3.24 \times 10^{-25}\,\mathrm{kg}$
\cite{shi2009molecular}.  The liquid-solid interaction is modelled by
the 12-6 LJ potential with $\epsilon_\mathrm{ls}/k_\mathrm{B} = 444\,$K and $\sigma_\mathrm{ls} = 0.28\,$nm to create a fully wettable substrate (zero contact angle).
The initial configurations of droplets are cut from a liquid bulk, created from equilibrium NVT simulations with a Nosé-Hoover thermostat at $T=400\,$K. The same ensemble and thermostat is used for the main simulations with the time step, 2.5 femtoseconds.
The entire domain width along the $z$-axis, $W=2\,$nm and the length scale for nondimensionalisation, $h_0=10\,$nm.

To compare MD results with the predictions of the SLE, the liquid transport properties are calculated are required. 
Here, dynamic viscosity is found to be $\mu = 1.64 \times 10^{-4}\,\mathrm{kg}\,\mathrm{m}^{-1}\,\mathrm{s}^{-1}$ by the Green Kubo method \citep{green1954markoff, kubo1957statistical}, which integrating the time-autocorrelation function of the off-diagonal elements of the pressure tensor $P_{ij}$ so that
\begin{equation}
\label{eq_Green_Kubo}
\mu = \frac{V_\mathrm{bulk}}{k_\mathrm{B} T} \int_0^\infty \left\langle P_{ij}(t)\cdot P_{ij}(0) \right\rangle dt \quad (i\neq j),
\end{equation}
where $V_\mathrm{bulk}$ is the volume of the bulk fluid, 
$k_\mathrm{B}$ is the Boltzmann constant, and $T$ is temperature. The pressure tensor components are obtained using the definition of \cite{kirkwood1949statistical} and the angular brackets indicate the expectation. 
The surface tension is calculated from the profiles of the components of the pressure tensor in a simple liquid-vapour system, using the mechanical definition \citep{trokhymchuk1999computer}:
\begin{equation}
\gamma = \frac{1}{2} \int_0^{L_z} \left[ P_\mathrm{n}(z)-P_\mathrm{t}(z) \right] dz,
\end{equation}
where $L_z$ is the length of the MD domain, and subscripts ‘n’ and ‘t’ denote normal and tangential components, respectively. 
Finally, we have $\gamma = 5.45 \times 10^{-2}\,\mathrm{N}\,\mathrm{m}^{-1}$.


\bibliography{apssamp}

\end{document}